\documentclass[iop]{emulateapj}

\shorttitle{Statistical Fitting}
\shortauthors{A. A. Breimann, S. P. Matt \& T. Naylor}

\usepackage{amsmath}
\usepackage[table,x11names]{xcolor}
\usepackage{ulem}
\usepackage{enumitem}

\usepackage{lineno}

\begin{document}


\title{Statistical fitting of evolution models to Rotation Rates of Sun-like stars}


\author{Angela A. Breimann*, Sean P. Matt \& Tim Naylor}
\affil{University of Exeter,
              Devon, Exeter, EX4 4QL, UK}
\email{*aab222@exeter.ac.uk}



\begin{abstract}
We apply for the first time a two-dimensional fitting statistic, $\tau^2$, to rotational evolution models (REMs) of stars (0.1 to 1.3 $M_{\odot}$) on the period-mass plane.  The $\tau^2$ statistic simultaneously considers all cluster rotation data to return a goodness of fit, allowing for data-driven improvement of REMs. We construct data sets for Upper Sco, the Pleiades and Praesepe, to which we tune our REMs. We use consistently determined stellar masses (calculated by matching $K_\textrm{s}$ magnitudes to isochrones) and literature rotation periods. As a first demonstration of the $\tau^2$ statistic, we find the best-fitting gyrochronology age for Praesepe, which is in good agreement with the literature. We then systematically vary three parameters which determine the dependence of our stellar wind torque law on Rossby number in the saturated and unsaturated regimes, and the location of the transition between the two. By minimising $\tau^2$, we find best-fit values for each parameter. These values vary slightly between clusters, mass determinations and initial conditions, highlighting the precision of $\tau^2$ and its potential for constraining REMs, gyrochronology, and understanding of stellar physics. Our resulting REMs, which implement the best-possible fitting form of a broken power-law torque, are statistically improved on previous REMs using similar formulations, but still do not simultaneously describe the observed rotation distributions of the lowest masses, which have both slow and fast rotators by the Praesepe age, and the shape of the converged sequence for higher masses. Further complexity in the REMs is thus required to accurately describe the data.
\end{abstract}





\section{Introduction}

The evolving rotation rates of low mass stars $(0.1 $ to $ 1.3 M_{\odot})$, which host convective envelopes and are magnetically active, inform us about the properties of stellar winds and can be a useful indication of stellar age \citep{Skumanich1972,Barnes2003,Meibom2015}. Given an initial age, rotation rate and mass, a star's rotational evolution can be described by taking into account 1) stellar structure, which initially spins the star up during pre-main sequence (PMS) contraction, and 2) stellar winds, which shed angular momentum and act to slow the star down over long (several tens to hundreds of Myr) time-scales \citep{WeberDavis1967,Kawaler1988}.

Observations of open clusters show a snapshot of how the rotation rates vary as a function of mass for a population of stars that are coeval. The youngest of these show an insight into the initial angular momentum distributions during the PMS, at which time significant changes in stellar structure and hence rotational evolution occur over just a few Myrs \citep{Irwin2009,Bouvier2014}. At these young ages, the presence of a circumstellar disk likely has a significant impact on the stellar rotational evolution. Accretion of disk material and contraction of the star act to spin the forming star faster, while magnetic interactions and accretion-related outflows remove angular momentum (see \citealt{Bouvier2014} for a review). PMS rotational evolution is a dramatic interplay of these three effects, and describing these phenomena remains a substantial challenge. Some rotational evolution models attempt to model this phase in detail \citep{Cameron1993,ArmitageClarke1996,Matt2012b}, but most treat this phase very simply (e.g. \citealt{GalletBouvier2013,GalletBouvier2015}), or initialise after some disk lifetime to avoid these complications (e.g. \citealt{Matt2015}). Modelling star-disk interactions, or implementing representative initial conditions into spin models is therefore of importance. 

Past the PMS stage, when the circumstellar disk no longer exists and stellar structure has settled, successively older clusters show a strong mass-dependent convergence of their rotation rates. The higher mass stars spin down quickly, leaving a dearth of fast rotators, and forming a `gap' between fast, higher mass rotators and slow, lower mass rotators \citep{Barnes2003}. The spin-down pattern shifts to lower masses systematically with cluster age. Once stars converge at slower rotation, they are observed to spin down approximately as $P_{rot} \propto t^{0.5}$ \citep{Skumanich1972}. The structure in the period-mass diagram during a cluster's initial stages therefore becomes erased, giving a direct relationship between age and rotation. This is the basis of gyrochronology \citep{Barnes2007}, a technique used to date stars. During the main sequence (MS), this technique complements other age indicators \citep{Angus2019}, such as lithium abundances or isochronal ages which can struggle to determine a precise age due to negligible changes in stellar structure. Rotation rates on the other hand can change over an order of magnitude during this time \citep{BarnesSpadaWeingrill2016}.  

Physics-based models of rotational evolution (or `spin') can currently broadly replicate observed spin distributions for clusters of varying ages (e.g. \citealt{Brown2014,Matt2015,Gondoin2017,Garraffo2018}), but none can precisely replicate all observed features in the period-mass diagram (PMD) across all ages. In this paper we focus on our own model based on \cite{Matt2015}, which describes broad features in the PMD, is able to reproduce the observed gap in rotation rates for young clusters, and forms a converged sequence of slow rotators. The shape of the converged sequence however does not exactly follow that of the data, and there is a population of stars at low mass and slow rotation that the model is unable to fit. 

It is largely the stellar wind that governs the rotational evolution of low-mass stars on the MS. Although there are some methods of inferring stellar wind properties, such as measuring Ly $\alpha$ emission \citep{Wood2005}, measuring stellar winds is difficult and very few measurements exist. Instead, the stellar wind torques used in spin models are largely informed by Magnetohydrodynamic (MHD) simulations, which describe how the torques vary as a function of stellar parameters such as magnetic field strength, mass loss rate due to stellar winds, stellar radius, mass and rotation rate \citep{Vidotto2011b,Matt2012,Finley2017}.

Additionally, internal angular momentum transport processes are likely to also have an effect on the rotational evolution, particularly for stars that are not fully convective (above approximately 0.35 $M_{\odot}$). These higher-mass stars have a radiative core and an outer convective layer and have been modelled as either a singular solid body (e.g. \citealt{Matt2015,vanSaders2019}); a double-zone model in which the core and envelope rotate as solid bodies whose angular velocities can differ \citep{MacGregorBrenner1991,Spada2011,GalletBouvier2013}; or as a multi-zonal structure \citep{Amard2016}. Angular momentum transport between different bodies has been modelled using a simplified coupling time-scale (e.g. \citealt{Irwin2007,Bouvier2008,Spada2011}) or can be self-consistently determined by including various transport processes in stellar interior models (as in \citealt{Amard2016}). 

The fact that no models to date fully predict the observed stellar densities of stars across the entire region of the PMD is an artefact of our current level of understanding of the stellar wind torques and internal angular momentum transport. There exists however a wealth of high precision rotation periods from space-based missions such as CoRoT \citep{Affer2012}, Kepler \citep{McQuillan2014}, and Gaia \citep{Lanzafame2018}, with many more expected in the near future. As a result, data-sets abundant in rotation periods are available for, to name but a few, clusters such as the Pleiades \citep{Rebull2016}, Praesepe \citep{Douglas2017}, Rho Oph and Upper Sco \citep{Rebull2018}. This flood of cluster rotation data, combined with a two-dimensional fitting statistic, is key to calibrating rotational evolution models. Whilst visually these models overlap strongly with the data, in order to improve on them there is a need for a quantitative fit.

The $\tau^2$ statistic, first developed by \cite{NaylorJeffries2006} in the context of fitting isochrones in the colour-magnitude plane, minimises when model probability densities on the two-dimensional plane optimally overlap with an observed dataset. In this paper, we adapt the $\tau^2$ data-driven statistical fitting technique to the period-mass plane using open cluster data, and implement it with our spin model. The goodness of fit is assessed by calculating the $\tau^2$ parameter, which is minimised for best-fit parameters in our torque law. 

A parameter study in the torque law of the models allow us to demonstrate the technique's ability to improve on the models, which will lead to a better understanding of stellar wind physics that lead to the observed mass-dependent distribution of rotation periods. We also demonstrate a technique which allows for the implementation of more sophisticated initial conditions into the rotational evolution models. In this paper, Section \ref{data_section} describes the three keystone cluster datasets (Upper Sco, the Pleiades and Praesepe) to which our rotational evolution models are fitted, and our method used to determine consistent masses across each. Our stellar wind torque law and the method for computing the rotational evolution for a single star is described in Section \ref{RotationalEvolutionModels}. Our $\tau^2$ technique requires the models to be expressed as a probability density distribution across the period-mass plane. Section \ref{GenSynthClust} describes how we initialise and evolve the rotation rates of synthetic clusters, and explains our method of obtaining a density distribution on the period-mass plane. In Section \ref{Tau2Main}, our $\tau^2$ fitting technique is outlined and demonstrated by finding a best-fit model dependent gyrochronology age for Praesepe.  In Section \ref{Results} we demonstrate the technique's versatility and potential to improve the models by finding best-fit parameters in the stellar wind torque law for (a) different clusters, (b) different mass transformations, and (c) different initial conditions. The resulting torque laws are presented and discussed in Section \ref{ResultingTorqueLaws}, and we conclude our work in Section \ref{Conclusion}. 

\section{Data}
\label{data_section}

Open clusters across a variety of ages are the ideal laboratories to constrain the rotation rates of a population of stars that are coeval. To improve the models in a data-driven manner, it is vital that the data sets we use to tune our models are a) rich in rotation periods and b) have masses determined consistently across all clusters. Otherwise, when finding best-fit values for free parameters in the model, it is possible that the parameters found will compensate for discrepancies in the mass determination methods rather than the physics of stellar winds. We therefore determine our masses consistently across all our cluster datasets.

We choose three key clusters that fulfil our criteria: Upper Sco, the Pleiades and Praesepe. Each of these have reliable distances, are close by, have good membership and recent measured K2 \citep{Howell2014} rotation periods. Their ages range from approximately 8 Myr to 665 Myr, which span the PMS and young MS. Combined, they give a distribution of period and mass over a key range of ages, which provides excellent constraints for the models.

Our theoretical models are expressed in terms of physical quantities such as mass and moment of inerita. In comparing model distributions to data it is therefore necessary to convert observable quantities such as colour and magnitude into mass. We choose to do so by using isochrones, which give a direct comparison of how colour (temperature) or magnitude (roughly luminosity) relates to the stellar mass.

Current stellar isochrones do not accurately describe the location of M stars on the colour-magnitude diagram (CMD), which is likely due to the underestimation of stellar radii by 10 percent (e.g. \citealt{Jeffries2017}). This is a problem both during the pre-main sequence \citep{Hartmann2003,Stauffer2007,Jeffries2017} and during the main-sequence \citep{Morrell2019}. When masses are inferred from their CMD location, they are particularly uncertain for masses $<$ 0.5 $M_{\odot}$, where masses can be under-predicted by as much as 20$\%$ \citep{KrausHillenbrand2007}. 

\cite{Bell2012} presented evidence that the $K_\textrm{s}$ band correctly predicted stellar masses. They thus posited that the $K_\textrm{s}$ magnitude correctly predicted the stellar mass, but that since the location of the model isochrone and the data is mismatched, the fault must largely be with the predicted optical colours. The optical colours in the isochrone were then adjusted until they predicted the same mass as the $K_\textrm{s}$ band. In this manner, the authors tuned unreliable optical colours such that the model better matched the observations, and made these semi-empirical isochrones available for use.

We used these tuned model isochrones for masses between 0.1 and 1.4 $M_{\odot}$ which consist of \cite{Baraffe1998} stellar interior models with mixing length $\alpha$=1.9, BT Settl atmospheres \citep{BTSettl} and a 2MASS system response from the online Cluster Collaboration Isochrone Server resource, hereafter named CCS \citep{Bell2014}. The advantage of CCS isochrones is that they are a better fit to the data, making them convenient to derive global cluster properties such as the age of Upper Sco and the reddening in optical colours for the Pleiades and Praesepe. However, we infer stellar masses for each cluster by matching observed $K_\textrm{s}$ magnitudes to those predicted by the isochrone, since we expect the $K_\textrm{s}$ band to be a good indication of stellar parameters \citep{Bell2012}. We emphasize that although we use semi-empirical tuned isochrones, the $K_\textrm{s}$ band is unaffected by any tuning, and thus the masses determined are also unaffected.

Finally, objects that have a colour-magnitude combination that is not described by the isochrone (which has $K_\textrm{s}$ magnitudes ranging from 8.08 to 14.17 and $(V-K_\textrm{s})$ colours between 1.06 and 6.50) are cut from the sample. We do this for the clusters Upper Sco, the Pleiades and Praesepe, and compile their corresponding rotation periods from the literature. 
 
Although the rotational evolution of a close binary system is likely to cause a different rotational evolution scenario than a single star system (e.g. \citealt{Patience2002,Meibom2007}), the numbers of these systems are small \cite{Meibom2007}. We therefore assume that any binaries in our datasets evolve as single stars. 
 \subsection{Upper Sco}

We use K2 rotation periods, de-extincted $K_\textrm{s}$ magnitudes and de-reddened ($V$-$K_\textrm{s}$) colours from \cite{Rebull2018} for Upper Sco. \cite{Rebull2018} compiled this photometry from the literature, where $K_\textrm{s}$ magnitudes are primarily from the 2MASS database, or from the Deep Infrared Southern Sky Survey (DENIS; \citealt{Foque1995}). Their $V$ band magnitudes are from a variety of databases (such as APASS; \citealt{Henden2016}) or are transformed from Gaia \citep{Gaia2016} or Pan-STARRS1 \citep{Chambers2016} colours.

The exact age of Upper Sco is still a matter of debate, but it likely has an age of between 3 and 10 Myr \citep{Rebull2018}. To determine which age to adopt when determining masses, we use a distance of 140 $pc$ \citep{Rebull2018} and generate several isochrones tuned in the $V$ band using the CCS, as these are a better fit to the data. We compare the isochrones to the location of the stars in the $K_\textrm{s}$ vs $V-K_\textrm{s}$ CMD and adopt an isochrone at an age of 7.6 Myr, whose $K_\textrm{s}$ magnitudes we then used to infer masses. 

\subsection{Pleiades}


We use 2MASS $K_\textrm{s}$ magnitudes, $(V-K_\textrm{s})_0$ colours and Kepler rotation periods listed in \cite{Rebull2016} to obtain a sample of Pleiades stars with good membership. Rebull 2016 find V band magnitudes from a number of sources in the literature \citep{Johnson1958,Landolt1979,Stauffer1987,Kamai2014,Prosser1991,Stauffer1998a} or infer V from other, close bands found in \cite{Bouy2013,Bouy2015} and \cite{Zacharias2015}. We generated a tuned isochrone using the CCS with an age of 135 Myr \citep{Bell2014} and a distance modulus of 5.63 \citep{Bell2012}. Following \cite{Bell2012}, a reddening of $E(B-V)$\,=\,0.04 based on the mean extinction $A_v$\,=\,0.12 (see \citealt{Stauffer1998}) was used.

\subsection{Praesepe}

For the open cluster Praesepe, we refer to \cite{Douglas2017} who compiled a list of adopted rotation periods, photometry, and masses. We use the adopted rotation periods listed in \cite{Douglas2017}, which are compiled from the literature \citep{ScholzEisloffel2007,Scholz2011,Agueros2011,Delorme2011,Douglas2014,Kovacs2014} or are their own measured K2 rotation periods. They also list $K_\textrm{s}$ magnitudes from 2MASS for these stars.

For consistency, our own masses are calculated from photometry provided by \citealt{Douglas2017}. A $K_\textrm{s}$ vs ($V-K_\textrm{s}$) CCS isochrone was generated with a reddening of 0.027 \citep{BJTaylor2006}, a distance modulus of 6.32 and an age of 665\,Myr \citep{Bell2014}, and the $K_\textrm{s}$ band was then used to determine masses. We consider only stars with masses $\leq 1.35 M_{\odot}$, as higher than this our rotational evolution models approach the Kraft Break \citep{Kraft1967}.

Although masses determined here and by \cite{Douglas2017} are obtained using $K_\textrm{s}$ magnitudes, the two methods of transforming observables to physical quantities differ, leading to a difference in the masses obtained. In this work, unless stated otherwise, we will use our masses obtained using the \cite{Bell2014} isochrones to be consistent across all three cluster datasets. However, having two different mass transforms for Praesepe gives us the opportunity to examine what difference in masses different assumptions yield. We compare these two datasets in Section \ref{tau2differentmasstransforms} of this paper. 

\cite{Douglas2014,Douglas2017} calculated masses for their entire sample using absolute $K_\textrm{s}$ magnitudes and relationships from \cite{KrausHillenbrand2007} which relate photometry and mass based on spectral energy distributions and additionally account for under predicted masses $<$ 0.5 $M_{\odot}$ by increasing the affected masses by a percentage based on how underpredicted they are. Although $K_\textrm{s}$ magnitudes seem a reliable indication of stellar properties \citep{Bell2012}, the stellar SED library used to estimate masses relates physical stellar properties with colours, which depend more strongly on the outer atmospheres of a star and are more uncertain. 

\begin{figure}[!htb]
\centering
     \includegraphics[trim=0.5cm 0cm 0.4cm 1.1cm, clip, width=1.1\linewidth]{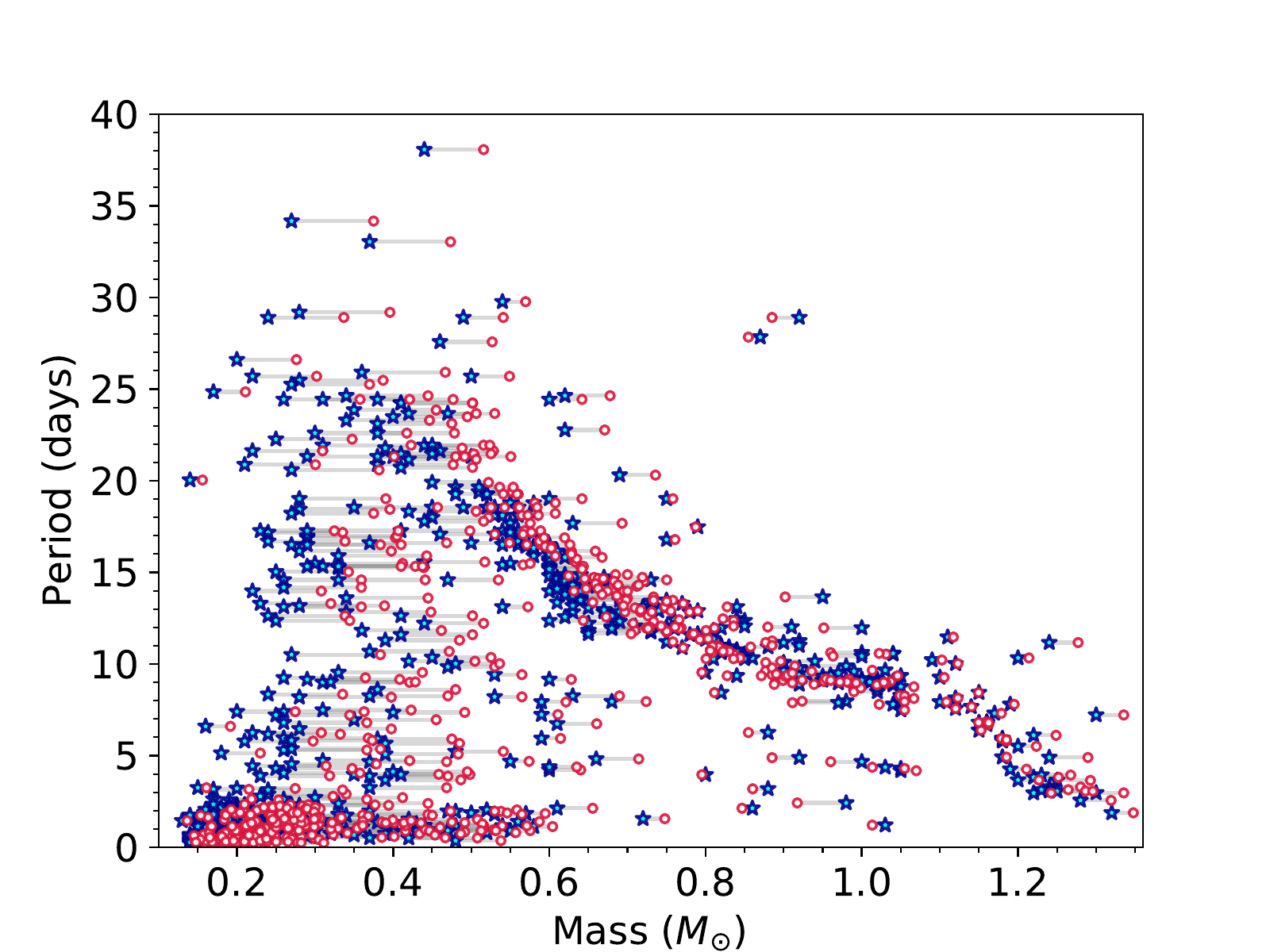}
     \caption{Praesepe rotation period versus mass data determined in this work (red circles) and in \citealt{Douglas2017} (blue stars). The difference in our method is most noticeably a shift to higher masses, with connecting grey lines demonstrating where each star is shifted according to the choice of mass transform.}
     \label{MassTransforms}
\end{figure}

The comparison of the two datasets is shown in Figure \ref{MassTransforms} which demonstrates the differences between mass transformation methods. The Figure shows that the discrepancies in mass are especially large at lower masses, where our method results in higher masses. The converged sequence, where the higher-mass stars with slower rotation rates have gathered together on the PMD, changes shape. The cut-off where the lower masses latch onto the converging sequence also changes location.

\section{The Rotational Evolution Models}
\label{RotationalEvolutionModels}
In this Section, we discuss how we generate models of stellar rotational evolution using the stellar wind torques of \cite{Matt2015}, hereafter referred to as the Classical Model, and our own wind torques. The latter reduces to the Classical Model for certain parameters. We therefore describe our torque law in this Section, and refer the reader to \cite{Matt2015} for further details.

\subsection{Formulation of stellar wind torques}

Our spin evolution model follows the prescription of \cite{Matt2015} where stars are assumed to rotate as solid bodies, and magnetic field strength $B_*$ and mass loss rate $\dot{M}_*$ scale with Rossby number according to a magnetic-activity function, which we name $F_\textrm{m}(Ro)$ and is described as

\begin{equation}
 F_{\textrm{m}}(Ro)= \textrm{Min} \bigg[ k_{\textrm{s}}  \bigg(   \frac{Ro}  {Ro_{\odot}}   \bigg )  ^{p_\textrm{s}} ,  \bigg(   \frac{Ro_{\odot}}  {Ro}   \bigg ) ^p   \bigg].
 \label{Fm}
\end{equation} 
The Rossby number
\begin{equation}
Ro = (\Omega_* \tau_\mathrm{cz})^{-1}
\label{Rossby}
\end{equation}
is an important parameter in stellar magnetism which relates the stellar rotation rate ($\Omega_*$) to the ratio of a typical time scale for convection in a convective region ($\tau_\mathrm{cz}$) \citep{Noyes1984}. The Rossby numbers in Equation \ref{Fm} have been normalised to the solar value, $Ro_{\odot}$. Here $p_\mathrm{s}$ is the scaling of the torque with Rossby number in the saturated regime, $p$ is the scaling of the torque in the unsaturated regime and the intercept of the saturated regime is $k_\mathrm{s}$. The first term in Equation \ref{Fm} represents the `saturated' regime wherein the magnetic activity saturates at a maximal value which has a weak dependence on $Ro$ (i.e. $p$ $>$ $p_\mathrm{s}$). The `unsaturated' regime is represented by the second term, where magnetic activity correlates with rotation for slower rotators. The point at which these power laws meet is the critical $Ro$, which varies with $p_\mathrm{s}$, $p$ and $k_\textrm{s}$. Our parameter search will be for values of $k_\textrm{s}$, $p_\mathrm{s}$ and $p$.

The torque for the entire star, which includes effects of stellar structure, is then 

\begin{equation}
T = T_{\odot} 
\bigg ( \frac{R_*}{R_{\odot}} \bigg) ^ {3.1}
\bigg ( \frac{M_*}{M_{\odot}} \bigg) ^ {0.5}
\bigg ( \frac{\Omega_*}{\Omega_{\odot}} \bigg) 
\beta ^ {-0.44} F_\mathrm{m} (Ro),
\label{torqueLaw}
\end{equation}
which is a function of the stellar radius $R_*$, mass $M_*$, and angular velocity $\Omega_*$, all normalised to the corresponding solar value (denoted with the subscript $\odot$). The constant $T_\odot$ is given by

\begin{equation}
T_{\odot} = - \bigg ( \frac{2}{p} \bigg) 6.3 \times 10^{30} \mathrm{\,\, erg}.
\label{eqn:torqueNormalisation}
\end{equation}

\noindent The factor in brackets shows the dependence of $T_{\odot}$ on the value of $p$ (the value of which governs stellar spin down), and ensures that the Sun is used as a constraint by requiring that solar mass stars spin down to the solar rotation rate by the solar age (see \citealt{Matt2015}). For $p$=2, $T_{\odot}$ reduces to the constant given in \citealt{Matt2015}.

The dimensionless number $\beta$, which aside from the case of the Classical model is always greater than unity, describes the influence of magneto-centrifugal acceleration on the speed of the stellar wind, and and takes the form

\begin{equation}
\beta = \bigg[1 + \bigg( \frac{f}{K_2}\bigg)^2 \bigg]^\frac{1}{2},
\label{beta}
\end{equation}
where $K_2$ is a constant determined by \cite{Matt2012} to be 0.0716, and where the stellar rotation rate is expressed as the fraction $f$ of the Keplarian rotation
\begin{equation}
f = \Omega_* \bigg( \frac{R_*^3}{GM_*}\bigg)^\frac{1}{2}.
\label{fractionVKep}
\end{equation}
The gravitational constant is donated by $G$. For $\beta$\,=\,1, $k_\mathrm{s}$\,=\,100, $p_\mathrm{s}$\,=\,0  and $p$\,=\,2, the torque law given by Equation \ref{torqueLaw} exactly reproduces that of the Classical Model. 

\subsection{Spin evolution calculations}
\label{SpinEvolCalc}
To calculate the rotation rate $\Omega_*$ as a function of time for a star, we solve the angular momentum Equation
\begin{equation}
\frac{\textrm{d} \Omega_*}{\textrm{d}t} = \frac{T}{I_*} - \frac{\Omega_*}{I_*} \frac{\textrm{d} I_*}{\textrm{d}t}
\label{angMomEqn}
\end{equation}
using a second order Runge Kutta numerical integrator with a variable time step and initial conditions of rotation rate, age, and mass. The method requires information about the stellar structure, including the moment of inertia $I_*$ and $\textrm{d}I_*/\textrm{d}t$ in Equation \ref{angMomEqn} as well as $M_*$ and $R_*$ in the torque law, which are obtained at each timestep from \cite{Baraffe2015} stellar structure models. Our models always use the same, solar, metallicity. The stellar wind torque $T$ is described in Equations \ref{Fm} - \ref{fractionVKep}. We thus generated spin tracks which describe the rotation rate of a star as a function of time, from the desired initial age to any final age reached in the stellar structure models. 

In this manner, spin tracks for stars of different masses, initial ages and rotation rates can be computed. In each case, the rotational evolution is as follows. We assume that at the chosen initial age of either 5 or 8 Myr, all stars have lost their disks. Despite the removal of angular momentum through stellar winds, the stars spin faster due to their contracting radii. Once these stars reach the MS, their stellar structure stabilises, contraction ceases, and the angular momentum evolution is instead governed by the magnetised stellar winds which remove angular momentum. The spin rates of these stars eventually converge in a mass dependent manner, where higher masses reach the converged sequence first, on the timescale of 100 Myr \citep{Bouvier2014}. 

\section{Generating synthetic clusters}
\label{GenSynthClust}
To generate an individual rotational evolution track, three initial conditions are required: a mass, an age, and an initial rotation rate. For a population of synthetic stars, several stars are initialised at the same age for a range of masses and initial rotation periods and evolved to some final time. The rotation rate of each star in the synthetic cluster at any age can be obtained from these spin tracks, and the model distribution at a given age on the period-mass plane can be generated. 

\begin{figure}[!htb]
\centering
     \includegraphics[trim=15. 1. 5. 30., clip, width=1.1\linewidth]{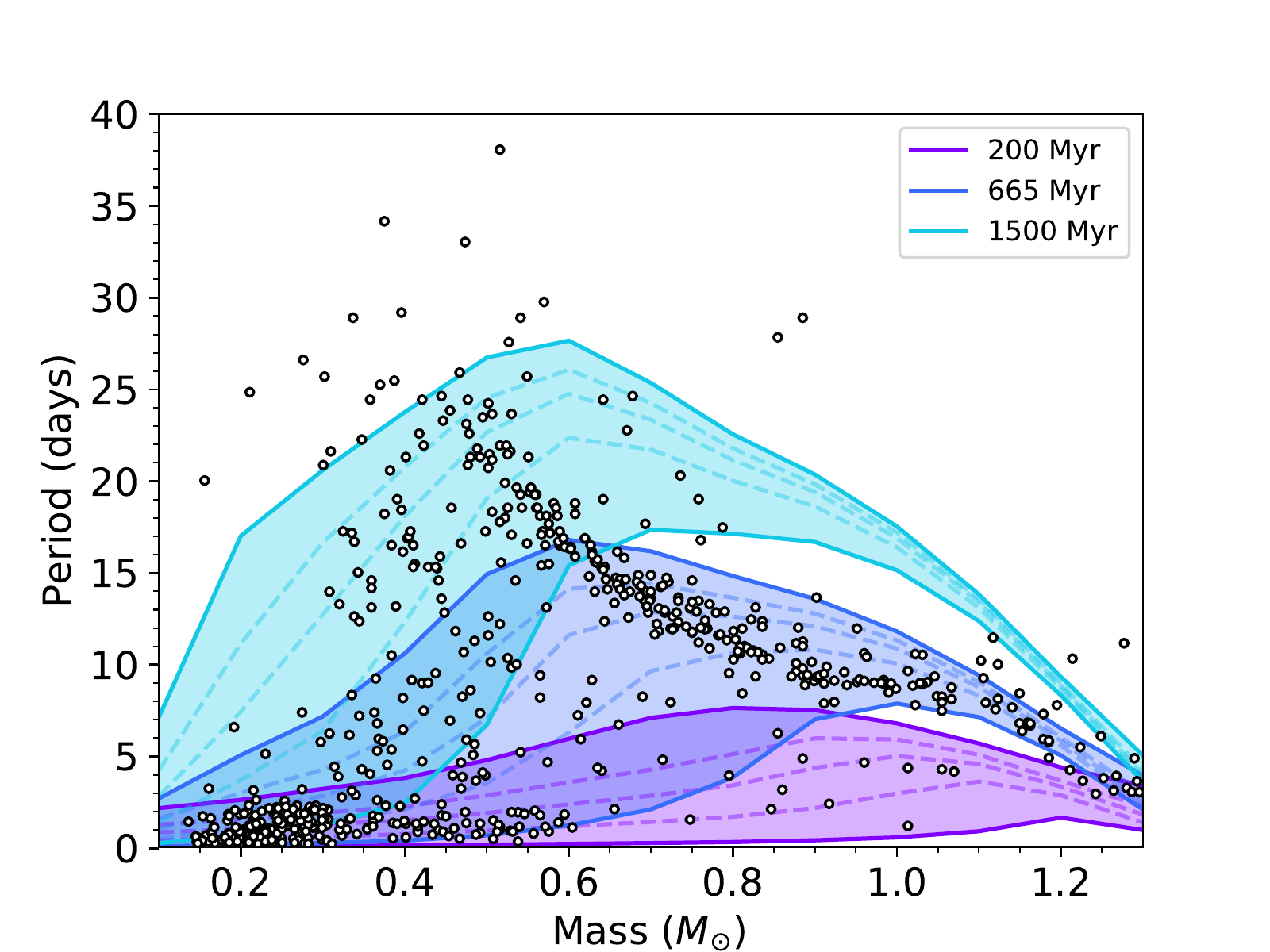}
     \caption{A period-mass diagram with Praesepe data (see Section \ref{data_section}) shown in black circles. The Classical model is shown at three different ages, demonstrating the evolution of the model spin distribution with time.  The youngest model age is at 200\,Myr (purple), followed by a model at 665\,Myr (blue) and the oldest at 1500\,Myr (turquoise). For each model age, the solid upper and lower boundaries represent the 0th (slowest stars) and the 100th percentile (fastest stars) of rotation respectively. At a given age, the model cannot explain any observed data anywhere outside this region enclosed by the solid lines. The dashed lines represent the 25th, 50th and 75th percentiles of rotation. Qualitatively, when these percentile contours are far apart, there is a smaller predicted density of stars, while when these lines gather, this shows a higher density. }\label{ModelEnvelopes}
\end{figure}

We create a grid of the Classical model consisting of 306 stars (initially evenly spaced in log period between 0.7 and 18 days) for each of our 13 masses ranging from 0.1 to 1.3 $M_{\odot}$. This initial distribution is hereafter referred to as the `Tophat' initial condition. An example of the resulting synthetic period-mass distributions is shown in Figure \ref{ModelEnvelopes}, which shows the representation of the Classical model evolved to three different ages, compared to the cluster Praesepe. Literature ages for this cluster range from 590 - 759 Myr \citep{Bell2014}, and throughout this paper we adopt a literature age of 665 Myr based on \cite{Bell2014}. The model ages shown are chosen to be at our adopted Praesepe age (665 Myr), too old (1500 Myr) or too young (200 Myr) to illustrate how the model changes with time. For a given model, the shape that represents it has an upper and lower solid line, corresponding to the very slowest and fastest rotators in the synthetic distribution. The 75th, 50th and 25th percentiles of rotation are also shown as dashed lines. When these lines are close together, the model predicts a high density of stars.

The models in Figure \ref{ModelEnvelopes} show the same general behaviour described in Section \ref{SpinEvolCalc} in the period-mass plane. The highest masses form a narrow sequence of rotation rates soonest, seen particularly well for the two oldest models of 1500 and 665 Myr, while at the same age lower masses remain much more spread in rotation rates. As expected, the model at 665 Myr overlaps with the bulk of the observed Praesepe cluster data. 

However, the model cannot reproduce the divergence of rotation rates at $\approx$ 0.4 $M_{\odot}$ where populations of both fast and slow rotators exist, failing to describe the slowest rotators entirely. Even where the model overlaps with the observed data, the highest predicted densities of stars are misplaced with respect to the data. This is particularly noticeable above 0.6 $M_{\odot}$, where the model predicts a high density of stars to be at longer rotation periods than is observed.

It is clear that, while the model does well at reproducing most features in the period-mass diagram, it still requires some tuning. A logical course of action is to express the model in terms of a density on the PMD. Doing so allows us to perform the $\tau^2$ statistic which gives the goodness of fit of a given model against a dataset in the period-mass plane. When changing model parameters to find a better model fit, $\tau^2$ allows us to determine which model is an improvement on the others. In Sections \ref{spinModelsAsProbDensity} and \ref{impInitCond}, we demonstrate how to construct a model probability density on the PMD using individual spin tracks, and in Section \ref{Tau2Main} we introduce and demonstrate the $\tau^2$ statistic.   

\subsection{Spin models as a Probability Density}
\label{spinModelsAsProbDensity}

There is a clear need to develop a statistical method to compare the spin model to observed period-mass data. This Section demonstrates how to express the model as a probability density on the period-mass plane.  This representation of the model allows for the prediction of the number of observed points in any given region of the PMD and hence the implementation of our $\tau^2$ goodness of fit parameter to discern a best-fit model.

A uniform population of a total of 3978 stars (306 stars for each of the 13 masses we have chosen) is first initialised as a Tophat distribution, in a manner similar to \cite{Matt2015}. This spin distribution is evolved to the desired age by computing spin tracks for each star. To give a synthetic probability density distribution on the period-mass plane at a given age, these model points are then binned into one of 13 mass bins 0.1 $M_\odot$ wide, starting at 0.05 $M_\odot$ and ranging to 1.35 $M_\odot$, and then further binned into one of 200 rotation period bins between $1 \times 10^{-5}$ and 40 days. This range in masses and rotation periods defines the area of our PMD. Unless stated otherwise, this setup remains constant throughout this paper.

In choosing an equal number of model stars in each mass bin, from which probability densities are calculated, we have chosen an equal weighting for each mass bin throughout this paper. This means our synthetic cluster always has a uniform initial mass function, which is explained in Section \ref{ModelNormalisation}.

The normalised model probability density $\rho(M,P)$ for each model bin at mass and period $(M,P)$ can be then calculated as the number of model points falling in that bin  $N(M,P)$ as a fraction of the total number of stars modelled $N_{tot}$. Finally, this fraction is divided through by the area of that bin $A(M,P)$, which has dimensions of $M_{\odot} \textrm{day}$.

Using too large a bin size results in resolution issues where density information within the model is lost, whereas too fine a bin size results in either empty or underpopulated bins. We adopt this binning approach throughout but ensure the model resolution is sufficiently high when using small bin sizes. 

The approach of populating a given bin with a model probability density described above is only valid if each model star has equal weighting. In Section \ref{UpperScoInitCond}, we demonstrate how allowing for different weightings gives the freedom to model different initial spin distributions.

\section{Tau Squared ($\tau^2$)}
\label{Tau2Main}

The $\tau^2$ maximum likelihood statistical fitting method was first introduced by \cite{NaylorJeffries2006} in the fitting of two-dimensional isochrone models to cluster data in colour-magnitude space. $\tau^2$ is a goodness of fit measurement that is minimised to find the most likely model. Our method is based on this work, applied for the first time in the context of rotational evolution models in period-mass space. Here we neglect the uncertainties in the mass and rotation period, the statistical uncertainty of which we assume to be negligible (or at least that the observed uncertainties are mostly contained within the relevant model grid cell for the majority of stars). 

\subsection{Intuitive interpretation of $\tau^2$}

When calculating the goodness of fit of a rotational evolution model to a cluster, we evolve the model to the desired cluster age and represent it as a surface probability density on the period-mass plane. To do this, the model is represented as a grid on the plane with each grid cell containing some model probability $\lambda(M,P)$, where $M$ and $P$ indicate the period-mass location of the grid cell within the grid. 

The model $\lambda(M,P)$ of each grid cell is normalised such that when all cell probabilities are summed, the total model probability across the period-mass plane is unity.

We define the surface probability density $\rho(M,P)$ in each grid cell such that the probability of the model $\lambda(M,P)$ within that region is
\begin{equation}
\label{lambdaDef}
\lambda(M,P)= \int_\mathrm{cell} \rho(M,P) \mathrm{\,d}M \mathrm{\,d}P,
\end{equation} 
where $\mathrm{d}M \mathrm{\,d}P$ represents the area in mass and period. 

Each observed cluster data point $i$ is then positioned on the period-mass diagram, where it is given an individual $\tau^2$ value based on the model probability density $\rho(M_i,P_i)$  at that location, (where $(M_i, P_i)$ is the model grid cell corresponding to the location of the $i^{th}$ data point), using 
\begin{equation}
\tau_i^2 = -2 \, \mathrm{ln} (\rho(M_i,P_i)).
\end{equation}
When points fall on regions of high model density, the overall contribution to $\tau^2$ will be small; and vice versa.

The goodness of fit of the rotational evolution model to a cluster's entire observed period-mass distribution is determined by calculating a total $\tau^2$ value, where the $\tau^2$ contributions of each data point are summed together such that 
\begin{equation}
\sum \tau^2_i = \tau_\mathrm{total}^2
\end{equation} 
This is the equivalent of multiplying several probabilities together. From this, a $\tau^2$ value associated with that data point can be calculated. The value of $\tau^2$ is minimised when the overlap between the 2-dimensional model distribution with the observed distribution is optimal.

\subsection{Derivation of $\tau^2$}

We base the following derivation of $\tau^2$ on the proof by \cite{Walmswell}, which describes the method in the case of a continuous model distribution, but does so by considering  infinitesimal regions of the period-mass plane.  The derivation also assumes no associated uncertainties on the measured data. The notation used throughout is listed in Table \ref{Tab:indices} for clarity.

\begin{table}
\caption{The notation and indices used for the proof of $\tau^2$.}
\begin{center}
	\begin{tabular}{ c|l } 
\hline
Index & Meaning    \\

\hline
i &  ith observed cluster data point \\
N &  total number of observed cluster data points \\
$P_i$ & period co-ordinate of the ith point  \\
$M_i$ & mass co-ordinate of the ith point \\
$\tau^2_i$ & $\tau^2$ of the ith data point \\
k &  number of data points expected to fall in a region \\
A &  number of regions containing one point \\
T &  total number of regions \\
a &  ath region containing one point \\
b &  bth region containing zero points \\
c & cth region containing either one or zero points \\
	\hline
	\end{tabular}
\end{center}
\label{Tab:indices}
\end{table}

For a model with probability densities $\rho(M,P)$, the probability $\lambda(M,P)$ of the model contained within a region in mass and period $\mathrm{d}M \mathrm{d}P$  is described by Equation \ref{lambdaDef}. To calculate the probability $P_{cell}(M,P)$ that some observed number of data points $k$ falls into a region at $(M,P)$ we can use the Poisson distribution 
\begin{equation}
P_{cell}(M,P) =  e^{-\lambda(M,P)}\frac{\lambda(M,P)^k}{k!}.
\end{equation}
If $\lambda(M,P)$ is small, $P(M,P)$ also becomes small as $k$ increases. In other words, for a model that predicts a low density of points in a given region, the probability of observing a high number of stars in this region is very low. We therefore consider the limit of small $k$, where we assume that either one ($k=1$) or zero ($k=0$) observed stars ever fall into a given region. 

To find the likelihood $L$ of the model given the data, the individual probabilities corresponding to each region (populated and unpopulated) are multiplied together, giving
\begin{equation}
L =\prod_{a=1}^{A} e^{-\lambda(M_a,P_a)}\lambda_{a}(M_a,P_a) \prod_{b=1}^{T-A} e^{-\lambda(M_b,P_b)} ,
\end{equation}
where $A$ is the number of bins occupied by one observed star, and $T$ is the total number of bins. The indices $a$ and $b$ correspond to the $a$th occupied region and the $b$th unoccupied region respectively. 

The $e^{-\lambda}$ terms can be gathered together into a sum that is over all pixels to give
\begin{equation}
L =\prod_{a=1}^{A} \lambda(M_a,P_a) \prod_{c=1}^{T} e^{-\lambda(M_b,P_b)},
\end{equation}
where the second product now has an index c, to represent that it is a product over all pixels. The second product can be written as a sum in the exponent

\begin{equation}
L = \textrm{exp}\bigg(- \sum_{c=1}^T \lambda(M_b,P_b)\bigg) \prod_{a=1}^{A} \lambda(M_a,P_a).
\end{equation}
The exponent is then reduced to 1 since since our model probability is normalised to unity over the whole plane, while $\lambda(M_a,P_a)$ is replaced by Equation \ref{lambdaDef} to give
\begin{equation}
L = [e^{-1}] \prod_{a=1}^{A} \rho(M_a,P_a) \mathrm{\,d}M \mathrm{d}P.
\end{equation}
Rearranging gives
\begin{equation}
L = \bigg[e^{-1}  \prod_{a=1}^A \mathrm{\,d}M \mathrm{d}P \bigg ]  \prod_{a=1}^{A} \rho_a(M_a,P_a),
\end{equation}
where the factor in the square brackets is a constant. Taking the log of this, describing the constant by $C$, and rearranging gives
\begin{equation}
 \mathrm{ln}L - C =  \sum_{a=1}^{A}\mathrm{ln} \rho(M_a,P_a).
\end{equation}
Multiplying both sides by $-2$ in analogy with $\chi^2$ \citep{NaylorJeffries2006} results in the general $\tau^2$ formula,

\begin{equation}
\tau^2_\mathrm{m} = -2(\mathrm{ln}L - C) = -2 \sum_{i=1}^{N} \mathrm{ln}  \rho(M_i,P_i),
\label{fulltau2}
\end{equation}
where the subscript $\mathrm{m}$ denotes the fit of the model to the data and where the sum is now over the number of stars in the dataset $N$ (which is equivalent to the number of populated regions for the finely sampled case). Here we have used small regions (where each model region contains either zero or one observed points) to derive the form of $\tau^2$. This breaks down for large region sizes, where a region can contain more than one star. In practice, for this paper, we approximate a finely sampled model with a more coarsely binned model. 

\subsubsection{Normalisation of model} 
\label{ModelNormalisation}

\cite{Naylor2009} show that requiring the model density to sum to unity across the entire plane is a convenient normalisation, which is enforced throughout this work. In addition, it is important to recognise that each stellar cluster has different numbers of stars with measured rotation periods and masses. It naturally follows that the cluster with the most stars would be weighted more strongly. If one wants to account for this and give each cluster equal weightings regardless of the number of stars they contain, the $\tau^2$ contribution from each data point $\tau^2/N$ can be calculated. In the work that follows, we do not implement this, and instead allow each star to be weighted the same, regardless of which cluster it is from.  

Observed clusters likely also have different initial mass functions and detection fractions of rotation periods, in which case areas of the period-mass diagram with more complete period detection could drive the fit. We do not account for this, since our aim throughout this work is to fit entire cluster datasets as they are observed and as a whole, without preferring any given feature across the PMD.

It is also important to note that individual $\tau^2$ values presented in this work carry little information, and change with resolution of model grid and of the initial conditions, as well as assumptions on the mass function applied to the synthetic cluster. Changing the latter adds a constant to the derived values of $\tau^2$ (see \citealt{Naylor2009}), and will therefore not affect best fit parameters. It is the \textit{relative} $\tau^2$ values that convey information about whether a given combination of parameters results in an improved model.

\subsubsection{Background probability values}
\label{normalisation}

The form of $\tau^2_\mathrm{m}$ (Equation \ref{fulltau2}) becomes problematic when a data point falls in a region of the period-mass plane at $(M_i, P_i)$ where the spin model does not exist ($\rho(M_i, P_i)$ = 0). To improve on the rotational evolution models, it is essential to include these outlying points in the $\tau^2$ analysis. This Section describes how we apply an additional probability across the entire plane, hereafter referred to as the `floor', in order to include these unmodelled points in our fit, as we ultimately want our model to describe all data.

To account for this, we follow the method described in \cite{Bell2013}, who, in the context of fitting stars to isochrones, describe a probability density of a population of contaminant non-member stars ($\rho_\mathrm{n}$) defined separately from a population of stars that are members of the cluster ($\rho_\mathrm{c}$), such that $\rho_\mathrm{n} +\rho_\mathrm{c} = 1$. Here we interpret the `contaminant stars' as the fraction of data points that the rotational evolution model is unable to describe.


In the context of the period-mass plane, it quickly becomes apparent that the normalisation of the rotational evolution model and that of the floor probability becomes important. As the total probability of the floor plus that of the rotational evolution model must always be unity, introducing a floor of probability across the period-mass plane requires (1) that an equal probability must be taken away from where the rotational evolution model exists, at the expense of the stars that the rotational evolution model does describe, and (2) that the period-mass plane must be a finite region sufficiently large so as to encapsulate both the model and all of the observed data. We define this region to be between 0.05 to 1.35 $M_{\odot}$ and $1 \times 10^{-5}$ and 40 days which is used throughout this paper. We add a constant floor probability density of $\rho_\mathrm{f}$ to all regions across this defined period-mass plane in addition to the model probability density $\rho$ to account for objects that are not described by the model. Each of these densities are then weighted by some value $\mathfrak{F}$ such that at any given point in the plane, the combined probability density $\rho'$  which includes both the spin model and the floor is 

\begin{equation}
\rho' = (1-\mathfrak{F})\rho_\mathrm{f}+\mathfrak{F}\rho.
\end{equation}
\noindent To find the value for $\rho_\mathrm{f}$, we follow the following reasoning. Since the total probability across the period-mass plane must be unity, the probability densities $\rho'$ integrated over the area of the plane must be equal to one, shown as

\begin{equation}
\int \bigg((1-\mathfrak{F})\rho_\mathrm{f}  +\mathfrak{F}\rho \bigg) \mathrm{d}A = 1.
\end{equation}
Since the model probability density $\rho$ is also normalised, the second term integrates to $\mathfrak{F}$. The first term is just an integral over $\mathrm{d}A$, since $(1-\mathfrak{F})\rho_\mathrm{f}$ is a constant. The integral therefore becomes

\begin{equation}
(1-\mathfrak{F})\,\rho_\mathrm{f} \, A_\mathrm{PM} +\mathfrak{F} = 1,
\end{equation}

\noindent where $A_\mathrm{PM}$ is a fixed rectangular region on the period-mass plane chosen to contain the full extent of the model and all observed data. This requires 

\begin{equation}
\rho_f = \frac{1}{A_\mathrm{PM}}.
\end{equation}
The total model probability density $\rho'$ is therefore of the form

\begin{equation}
\rho' = \frac{1-\mathfrak{F}}{A_\mathrm{PM}} + \mathfrak{F} \rho, 
\end{equation}
\noindent where we have defined the floor in terms of $A_\mathrm{PM}$. There is hence always some probability that an observed point is not a part of the spin evolution model. Including these outlying points in this manner ensures the spin model is penalised for the points it is unable to describe. The final form of $\tau^2$,

\begin{equation}
\label{tau2withfloor}
\begin{split}
\tau^2 = -2 \, \mathrm{ln} \sum_{i=1}^N    \rho'(M,P)\\
= -2 \, \mathrm{ln} \sum_{i=1}^N  \bigg( \frac{1-\mathfrak{F}}{A_\mathrm{PM}} + \mathfrak{F}\, \rho(M,P) \bigg),
\end{split}
\end{equation}
includes both the floor and the spin model probabilities in the fit, and is not to be confused with previously shown forms of $\tau^2$. It is this form of $\tau^2$ that we use for all our analysis throughout this paper.  We reiterate that both $\rho'$ and $\rho$ must be normalised such that $\sum{^N}_{i=1}\, \lambda'(M_i,P_i) = 1$ and $\sum{^N}_{i=1}\, \lambda(M_i,P_i) = 1$.

\subsubsection{Finding a value for $\mathfrak{F}$}
\label{findingF}

As discussed in Section \ref{normalisation}, there is a need for some floor background probability across the entire period-mass plane to account for observed points that fall outside where the spin evolution model exists ($\rho$ is 0). The value of this floor probability determines how much of a penalty is given in the $\tau^2$ fit for stars that are not described by the model. Too low a floor value causes the $\tau^2$ minimisation to find a model which describes most of the data, with little regards to how probable a point encapsulated by the model is. Conversely, too high a floor results in a model which best-fits the substructure of the model, without little penalty given to points completely outside the range of the model.

To find the appropriate value for the floor, we followed the prescription by \cite{Bell2015} to set $\mathfrak{F}$ to 0.7, having tested cumulative distributions of $\tau^2$ in a method similar to $\chi^2$ clipping. We found that our results in Section \ref{Results} were unaffected as long as $\mathfrak{F}$ was in the range of 0.6 to 0.9.

A $\mathfrak{F}$ of 0.7 led to the most self-consistent best-fit parameters. We adopt this value for the remainder of the paper. We also note that the value of $\mathfrak{F}$ (and hence the floor) will change the $\tau^2$ value, so it is important to compare relative rather than absolute values of $\tau^2$.

\subsection{Implementing $\tau^2$}
\label{Implementingtau2}

Because spin distributions have a wide dynamic range, we perform $\tau^2$ fits in both log and linear period space, which will be discussed in more detail in Section \ref{testingTau2}. When calculating $\tau^2$ fits in log period space, the grid is set up such that the mass binning remains the same but for the period dimension, the grid spacing is defined in log period space. The procedure is then essentially the same as Section \ref{spinModelsAsProbDensity}, where the calculation of model probability densities is described, and where the bin area $A(M_i,P_i)$ is calculated in the same way as in the linear grid spacing case. This leads to different model densities $\rho$ between the two spaces, and different values of $\tau^2$. 

In this paper, we hence calculate $\tau^2$ in both log and linear period space, with the spin model probability $\mathfrak{F}$ set to 0.7, following Section \ref{findingF}. To find a the goodness of fit of the resulting model with a given observed dataset, the individual $\tau^2$ contributions for each point in an observed dataset are summed using Equation \ref{tau2withfloor}.

\subsection{Demonstrating $\tau^2$ with Gyrochronology}
\label{testingTau2}

To illustrate the $\tau^2$ technique in a real case, in this Section we demonstrate that minimising $\tau^2$ gives a best-fit model age for Praesepe. In other words, we use $\tau^2$ to find a gyrochronology age. As the rotational evolution model used to obtain this age has already been visually tuned to the literature age of the cluster by the authors of \cite{Matt2015}, the age found in this exercise is not an independent determination and instead acts purely to demonstrate $\tau^2$.

\begin{figure*}
\centering 
\includegraphics[trim=2cm 0.1cm 0cm 1cm,clip,width=0.9\linewidth]{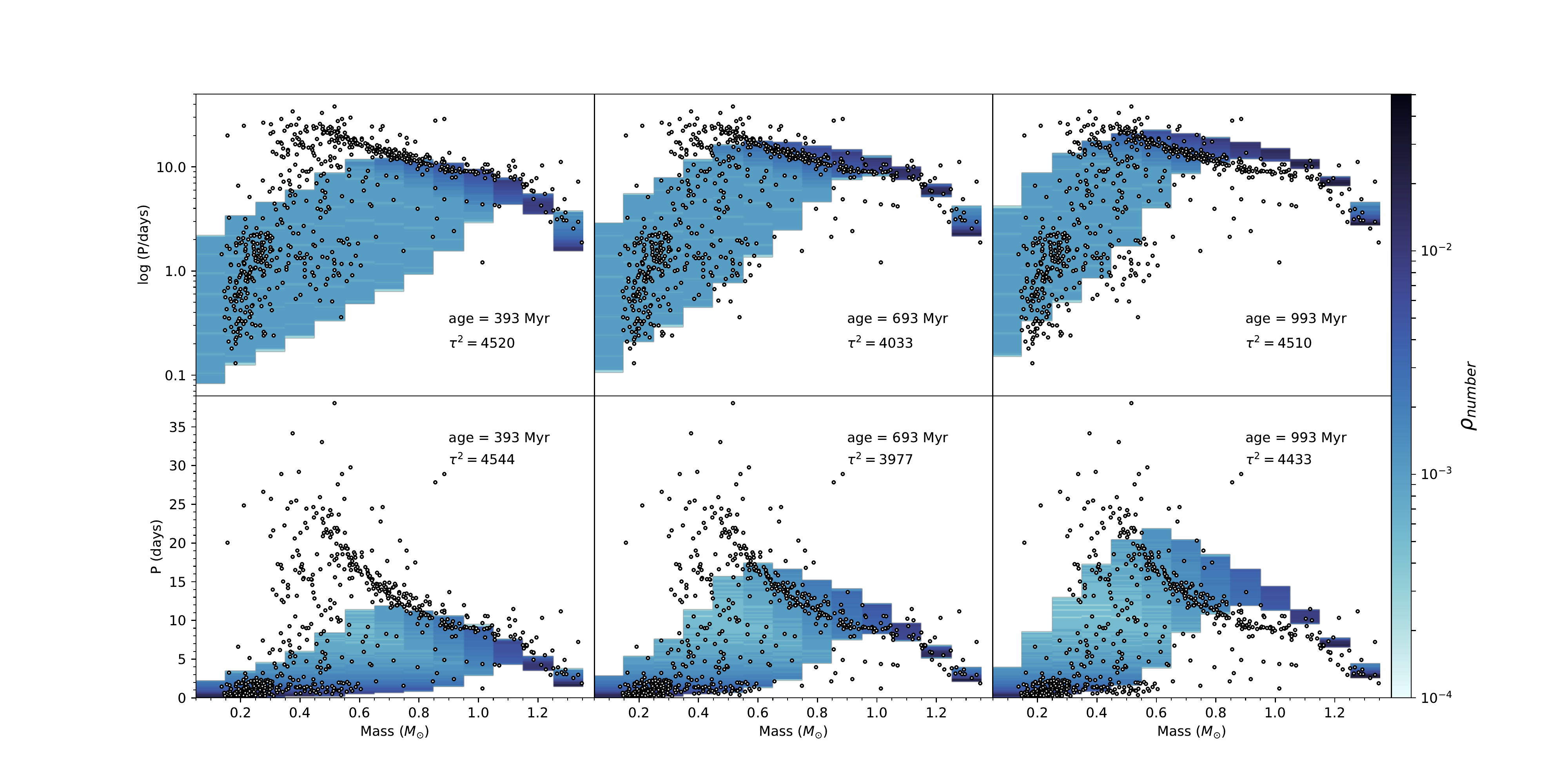}
\caption{Period-mass diagrams of Praesepe (black circles) with a log period scale (top row) and linear period scale (bottom row), compared to the classical model at three different ages of 410\,Myr (first column), 710\,Myr (middle column) and 1100\,Myr (final column). The model number density $\rho_\mathrm{number}$ of the classical model is plotted, where darker regions of blue denote where the model predicts a higher density of stars (note that $\rho_\mathrm{number}$\,=\,$A_{ij}\rho'$). Each panel shows the age and goodness of fit statistic $\tau^2$ value of each fit. The best-fit model is shown in the middle, with a model that is too young shown to the left and too old shown to the right.}
\label{fitsToPraesepe}
\end{figure*}

Following the method in Section \ref{GenSynthClust}, we evolved an initial synthetic distribution of stars according to the Classical spin model. From these, we generated model probability density grids assuming Tophat initial conditions in both linear and log period space. These probability density grids were obtained for ages at every Myr between 5 and 2000 Myr using a high resolution binning of 200 bins in the period dimension.

For each model age in our grid, we performed a $\tau^2$ fit against Praesepe to obtain $\tau^2$ as a function of age. Figure \ref{fitsToPraesepe} shows Praesepe at three of these ages to demonstrate how the model evolves with age. Log and linear period spaces are shown as they each highlight underlying density structure in different regions. Linear space is particularly sensitive to slow rotators (and hence the sequence at higher masses), whereas log space focuses on the distribution of the fast rotators (mostly at lower masses). When comparing the model to a dataset, it is important to first assess the overlap of the model shape with the data. Secondly, one can also look for structure within the data that is repeated by the model. 

The left-most column of Figure \ref{fitsToPraesepe} shows a model that, at an age of 393 Myr, is too young to describe Praesepe. Despite describing the clump of faster rotators at low mass well (as seen in log period space), the model barely describes the bulk of the slowest rotators at the highest masses, missing a large fraction of data points entirely at around 0.5 $M_{\odot}$ (particularly noticeable in linear period space). The right-most panel on the other hand shows a model at 993 Myr, which is much older than Praesepe. The upper log plot shows that the model has begun to evolve past the clump of fast rotators at low mass. Its converged slow-rotator sequence at the highest masses also largely misses the data. The middle panel shows the best-fit model at an age of 693 Myr. The clump of fast rotators at low masses are largely described. The model has mostly evolved past the slow-rotating high-mass converged sequence, but the model still describes more of the faster rotators, which is detected by $\tau^2$. There is therefore a play-off between the converged sequence and the lower masses as to which fits better in each space.

Even for the best-fit age (middle column of Figure \ref{fitsToPraesepe}),  it is apparent that there are regions in the period-mass diagram where the Classical model consistently does not reproduce the observations.  In particular, the triangular region of slowly rotating stars at $\sim0.5\,M_{\odot}$ is never described well by the model. In addition, the model converged sequence of slowly-rotating high-mass stars, while largely overlapping with the data, does not follow the shape of the sequence formed by the data. It is these regions that we aim to improve the fit in future generations of spin models.

\begin{figure}[h]
\centering 
\includegraphics[width=\hsize]{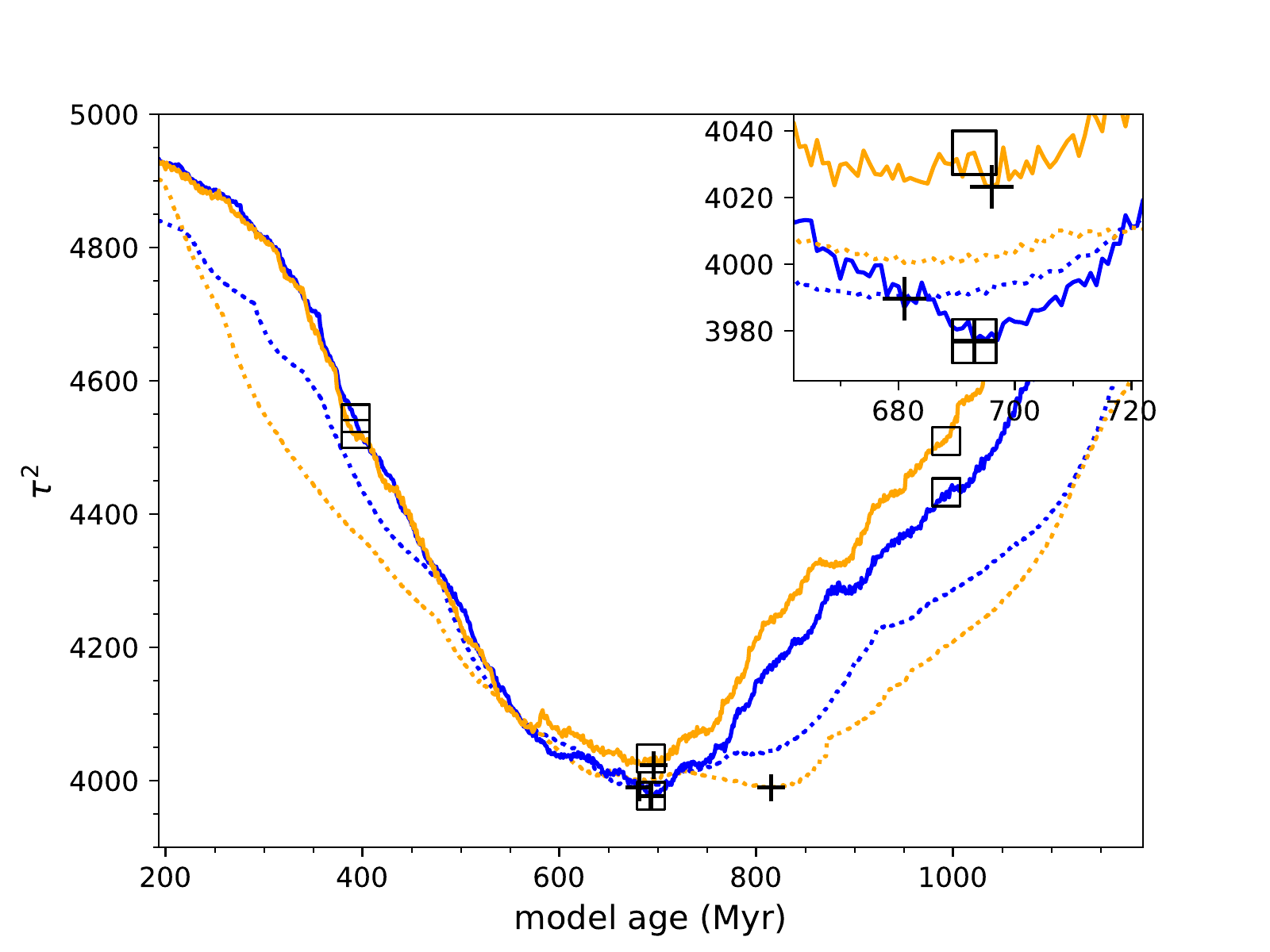}
\caption{The $\tau^2$ fit of the spin model to Praesepe as a function of model age is shown for different model grid setups. The minimum value of $\tau^2$ found in each case corresponds to the best-fit age. The high resolution runs (200 bins in period space) are shown by the solid lines, and low resolution (15 bins) by the dotted lines. Blue corresponds to $\tau^2$ calculated in linear period, and orange in log period space. The location of the best-fit age in each case is shown by the cross, while open squares show the locations of the model distributions in Figure \ref{fitsToPraesepe}. The inset panel shows a closer view of three of the best-fit ages. The figure demonstrates that best-fit values which closely agree in both log and linear period space are achieved for a sufficiently high model resolution grid in the period - mass plane.}

\label{gyroPlot}
\end{figure}

Figure \ref{gyroPlot} shows the $\tau^2$ obtained as a function of age for fits in both log and linear rotation period space for fine (200 bins) and coarse (15 bins) binning in period space respectively. The ages for which $\tau^2$ is minimised in each of the four cases is shown, as are the ages and $\tau^2$ values corresponding to Figure \ref{fitsToPraesepe}. 

With high resolution binning, the best-fit ages found in linear space of 693\,Myr ($\tau^2$\,=\,3977) and 696\,Myr in log space ($\tau^2$\,=\,4023) closely agree within just a few Myr. The low resolution run in linear space finds a similar best-fit age of 681\,Myr ($\tau^2$\,=\,3990). While the corresponding log space run finds a different age of 815\,Myr ($\tau^2$\,=\,3990) in a global minimum, there is a local minimum occurring at a very similar age to the other runs. This mismatch in best-fit ages found in the lower resolution run indicates that the model is not sufficiently resolved on the period-mass plane and leads to disagreeing best-fit parameters between the two spaces. We therefore adopt the high resolution binning of model densities on the PMD for the rest of this paper.

Aside from one inconsistent result, the best-fit ages found by $\tau^2$ are in broad agreement with literature ages, which range from 590 to 800 Myr \citep{Fossati2008,Brandt2015}. Our age of 693\,Myr is in particularly good agreement with the isochronal age of $665^{+14}_{-7}$\,Myr found by \cite{Bell2014}. While this is unsurprising - the spin models have been tuned to cluster rotation datasets - it shows statistically that the models fit the bulk of the observed distribution despite their flaws, while also demonstrating that our $\tau^2$ statistic can be used for gyrochronology.

\section{Using $\tau^2$ to tune Rotational Evolution Models}
\label{Results}

We next implemented $\tau^2$ to demonstrate its ability to find best-fit parameters in our stellar wind torque law (Equation \ref{torqueLaw}) with the goal of improving the current fits of the models to the data as shown in Figure \ref{fitsToPraesepe}. We do this for three situations; first we demonstrate $\tau^2$ applied to the Pleiades and Praesepe and its sensitivity to the different clusters; then we test its sensitivity to different mass transforms; and finally we show its ability to determine which initial conditions are preferable.

Our method is similar to that in Section \ref{testingTau2}, but instead of varying just the model age, we instead fix the model age to our adopted literature cluster age and vary three parameters in Equation \ref{Fm} of our torque law, $k_\textrm{s}$, $p_\mathrm{s}$ and $p$. These parameters represent how the stellar wind torque of these stars scales with Rossby number, and we choose these because the Rossby dependence of torques is uncertain and has been the key `tunable' parameter in many previous works. The slope and level of the saturated regime are represented by $p_\mathrm{s}$ and $k_\textrm{s}$ respectively, while $p$ describes the slope of the unsaturated regime. We tie the normalisation in the torque to the value of $p$ (Equation \ref{eqn:torqueNormalisation}) to ensure the solar-mass stars reach approximately solar rotation at the solar age. Each combination of these parameters additionally gives rise to a different value for the critical Rossby number whereby a star enters from one regime to the other (see \citealt{Matt2015}). By varying these parameters we determine how long these stars stay in the saturated regime, during which time their spin distributions retain a broad distribution, and when they reach the unsaturated regime, when the spin rates converge. 

We implement a grid search `brute force' method to find the parameters that minimise $\tau^2$. The parameters range from $p$\,=\,1.7 to 2.5 in steps of 0.1, $p_\mathrm{s}$\,=\,$-$0.4 to 0.5 in steps of 0.1, and $k_\textrm{s}$\,=\,$-$200 to 1200 in steps of 100. For each combination of these parameters, we generate a set of spin tracks according to the new torque law, and make a model density distribution on the period-mass plane at the desired age. We do this for both linear and log rotation period versus mass, calculating the $\tau^2$ values for each space. We build up the 3-dimensional parameter space in this manner. 

For each of the following three Subsections, we calculate a 3-dimensional $\tau^2$ grid for both the Pleiades and Praesepe. We also calculate the $\tau^2$ grid that represents the goodness of fit of the model to both clusters simultaneously for each set of parameters. As $\tau^2$ is the log of the likelihood, the $\tau^2$ values for each cluster can be added to obtain the total $\tau^2$. This allows us to optimise the model that best-fits both clusters simultaneously, acting as a statistical compromise between the best-fits of the two clusters individually. 

It is important to note that this parameter space spans a wider range than we expect for natural stellar behaviour. A $k_\textrm{s}$ value of zero implies a stellar wind which carries away no angular momentum, and a negative $k_\textrm{s}$ means that stars gain angular momentum. Our results in each case show, as expected, a steep decline of probability for this parameter space.

We applied a prior in order to account for the unnatural parameter space described, but this did not change our results. It was therefore not accounted for in the calculations of this paper.  

\subsection{$\tau^2$ for different clusters}
\label{Matt2020Tau2Search}

\begin{figure*}
\centering 
\includegraphics[trim=2cm 0.1cm 0cm 1cm,clip,width=0.9\linewidth]{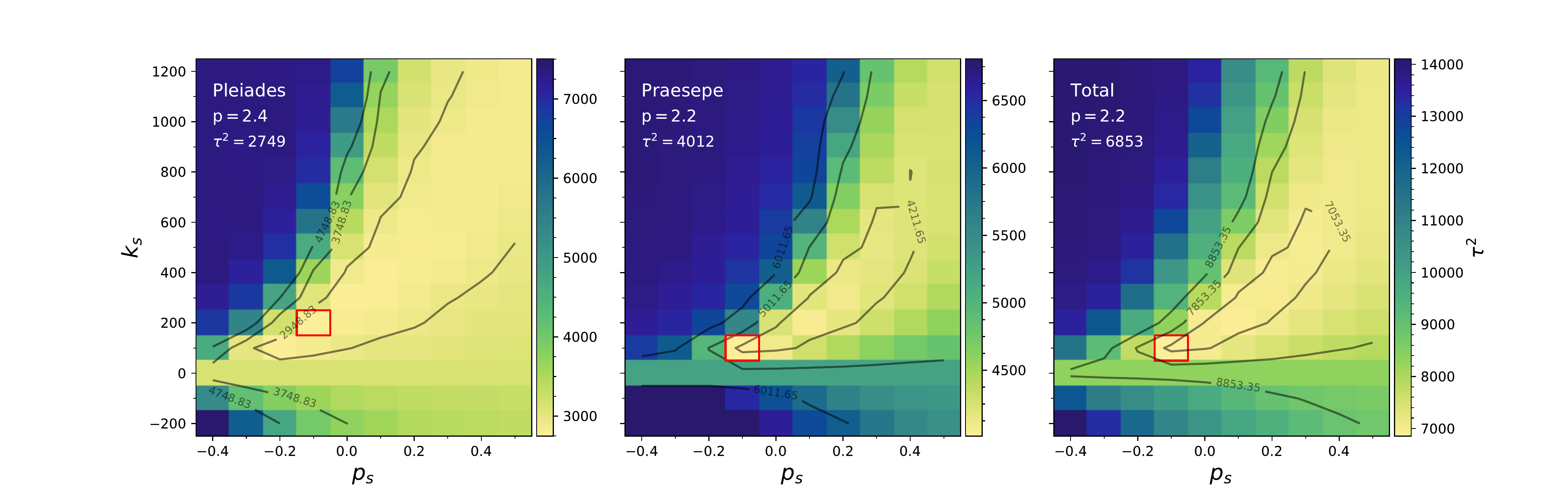}
\caption{The $\tau^2$ surfaces for $k_\textrm{s}$ vs $p_\mathrm{s}$ for the Pleiades (left), Praesepe (middle) and combined (right). Each panel shows a slice of the 3D parameter space (performed in linear period-mass) which corresponds to the best-fit $p$ value as indicated. Our torque law has been varied and the resulting spin model with Tophat initial conditions fit to the datasets determined in this work. Each pixel corresponds to the linear period-mass diagram obtained by evolving a spin evolution model to the appropriate cluster age for that combination of $k_\textrm{s}$, $p_\mathrm{s}$ and $p$ values, and the colour bar represents the corresponding value of $\tau^2$. The best-fit parameters correspond to the minimum value of $\tau^2$, which indicated by the red squares and are also listed in Table \ref{Tab:Matt2020} along with the corresponding results in log period space. Finally, contours of 200, 1000 and 2000 above the minimum value in $\tau^2$ space are shown to facilitate comparison between each surface. The difference between each $\tau^2$ minimum and the contours on the surfaces are equal steps in probability.}
\label{bestFitSurfaces}
\end{figure*}

\begin{figure*}
\centering 
\includegraphics[trim=2cm 0.1cm 0cm 1cm,clip,width=0.9\linewidth]{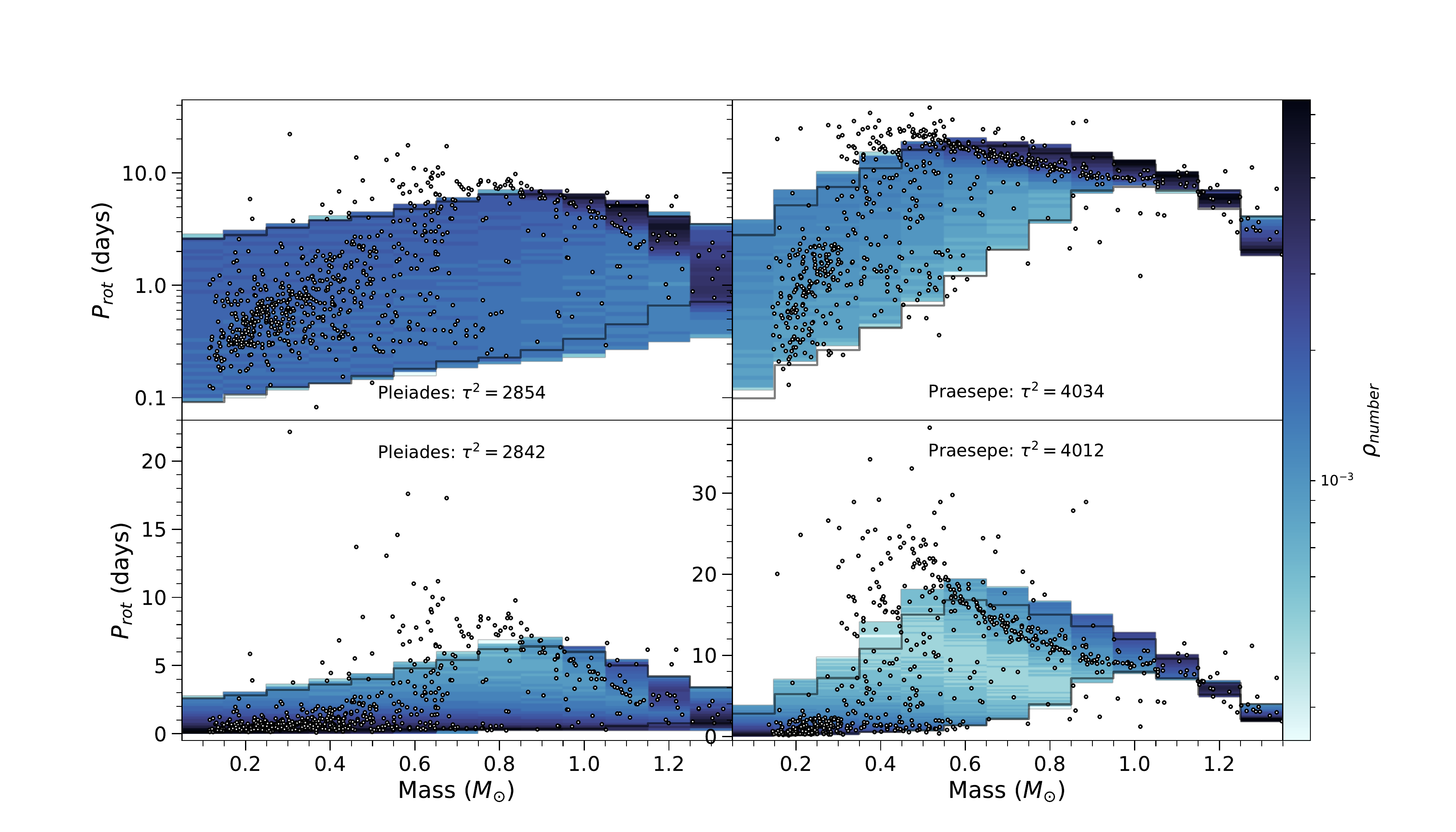}
\caption{Period-mass diagrams showing observed cluster data (black circles) with a log period scale (top row) and linear period scale (bottom row), compared to our model with best-fit values from the linear total $\tau^2$ surface shown in Figure \ref{bestFitSurfaces} ($p_{\mathrm{s}}$\,=\,$-$0.1, $k_\mathrm{s}$\,=\,100 and $p$\,=\,2.2). The first column shows the Pleiades cluster with the model evolved to 135 Myr, and the second column shows Praesepe with the model evolved to an age of 665 Myr. The model number density $\rho_{\mathrm{number}}$ of the classical model is plotted, where darker regions of blue denote where the model predicts a higher density of stars (note that $\rho_{\mathrm{number}}$\,=\,$A_{ij}\rho'$). Each panel shows the age and $\tau^2$ goodness of fit value, as well as the outline of the Classical model evolved to the appropriate age in grey.} 
\label{bestFits}
\end{figure*}

In this Section, the best-fit wind torque parameters for the Pleiades and Praesepe are found independently. We  run the models to literature ages of 135 Myr for the Pleiades \citep{Bell2014} and 665 Myr for Praesepe \citep{Bell2014} and generate model period-mass probability densities at these ages for each set of parameters. We then perform three-dimensional $\tau^2$ grid searches for the parameters $k_\textrm{s}$, $p_\mathrm{s}$ and $k$ on each cluster and the combined $\tau^2$ grid. 

The resulting grids are shown in Figure \ref{bestFitSurfaces}, where 2-dimensional slices of the 3-dimensional $\tau^2$ grid searches are shown for the Pleiades, Praesepe and the total $\tau^2$. These slices correspond to the best-fit $p$ value found in each case. Only the linear-space $\tau^2$ surfaces are shown, as they are very similar in appearance to the log-space surfaces. Each $\tau^2$ surface shows the same general trend, where there is a low $\tau^2$ (high probability) valley that depends on $k_\textrm{s}$ and $p_\mathrm{s}$, showing that as the slope $p_\mathrm{s}$ in the saturated regime is increased, the level $k_\textrm{s}$ is brought up to compensate. These two parameters are partially degenerate, and the valley follows the location of nearly constant critical Rossby number $\chi$ used in the \cite{Matt2015} torque law. Furthermore, each surface shows a steep increase in $\tau^2$ corresponding to a decline in likelihood for the unnatural parameter space of $k_\textrm{s}$ $\leq$ 0.

Despite the broad similarities of these surfaces, the low $\tau^2$ valley for the Pleiades and Praesepe differ slightly in both location and steepness. Whilst the Pleiades has a steeper slope away from the minimum in the $k_\textrm{s}$ (shown by closer contour lines), its highest confidence contour plateaus over a much larger range of parameter space, indicating that $\tau^2$ has been determined with less confidence for the Pleiades than for Praesepe. The total $\tau^2$ surface is a compromise between the two clusters, showing a steeper rise in $\tau^2$ from the minimum than shown in the case of Praesepe, with a much smaller area contained within the highest confidence interval than the Pleiades. 

The best-fit parameters and minimised $\tau^2$ values for each 3-dimensional parameter search are shown in Table \ref{Tab:Matt2020}. Despite a slight variation of these parameters between clusters, it is remarkable how close the values found are. In a broad sense, the same model is able to describe both clusters. Indeed, the parameters found in each case are similar to those of the Classical model, which has $p_\mathrm{s}$\,=\,0, $k_{\mathrm{s}}$\,=\,100 and $p$\,=\,2. The slight differences in parameters show how sensitive $\tau^2$ is to subtle variations between datasets.

The best-fit parameters for the Pleiades calculated in log and linear period space closely agree within the spacing of our parameters (i.e. they differ at most by 1 pixel in the grid). For Praesepe however, the $p_\mathrm{s}$ values vary by 2 pixels between the two spaces, which is likely a resolution issue of the spin model density on the log period-mass plane (as discussed in Section \ref{testingTau2}). Since the Pleiades and Praesepe also give different preferred torque parameters, we implement values found by the total linear $\tau^2$ surface to generate new best-fit spin models on the period-mass plane. These are shown in Figure \ref{bestFits} at the Pleiades and Praesepe ages in both log and linear period space, with the extent of the corresponding Classical model depicted in grey.

The distribution of our statistically-fit model has changed from the Classical model to encapsulate a greater number of observed points, particularly at the age of Praesepe (right column of Figure \ref{bestFits}). At this age, the upper model envelope has been pushed up in the PMD to include more slower rotators, while the lower envelope describes a similar number of faster rotators. Whilst the differences in torque parameters found for each cluster is a demonstration of the sensitivity of $\tau^2$, the resulting distribution of the spin models in the PMD remains very similar to that of the Classical model. Although our model describes more observed stars, there is still a significant population of low-mass slow-rotators that it is unable to describe. Additionally, it is unable to reproduce the distribution of observed data at higher masses, predicting the converged sequence to be at slower rotation than is observed.  The current torque law lacks sophistication to describe the physics of these stars. 

The $\tau^2$ results in Table \ref{Tab:Matt2020} indicate different clusters give very slightly different best-fit parameters. Comparing the linear-space results shows that the variation is primarily in $p$. The Pleiades require a slightly higher value ($p$=2.4) than Praesepe ($p$=2.2), in order that more stars spin down faster once they reach the unsaturated regime. This mismatch could be due to intrinsic differences within the clusters, such as environmental differences including proximity to disk-evaporating O stars \citep{Roquette2017}, or differing stellar metallicity \citep{Amard2019,Amard2020}. It could otherwise be due to the current fit of the models to the data. An improved spin model might find best-fit values that describe both clusters simultaneously using the same parameters with the same set of parameters.

We also notably find that Praesepe provides a much more stringent (and hence better) constraint on the spin models than the Pleiades. Although the Pleiades has more data-points than Praesepe (731 and 664, respectively), Praesepe consistently gives a higher $\tau^2$ value, indicating that our models fit Praesepe more poorly than the Pleiades. The difference in $\tau^2$ is much greater from pixel to pixel in the Praesepe surface than in the Pleiades surface. The effect of Praesepe's steeper gradient in $\tau^2$ parameter space is that Praesepe leads the fit of the total $\tau^2$ surface more strongly than the Pleiades. This is partially due to its longer lever arm in time: both the cluster and the model have had much more time to develop the appropriate structure at the Praesepe age than the Pleiades age, and  Praesepe also spans the saturated and unsaturated regime well, whereas the Pleiades does not. Re-weighting each of the data points in each cluster to account for differing numbers, following the methodology discussed in Section \ref{ModelNormalisation}, would weight Praesepe even stronger.

\begin{table}[h]
\caption{The best-fit $\tau^2$ and associated best-fit  torque parameters of our model obtained in both log and linear rotation period space for the Pleiades and Praesepe, as well as the total $\tau^2$. The parameters in bold have been used to generate the model distribution in Figure \ref{bestFits}.
}
\begin{center}
	\begin{tabular}{ l|c|c|c|c|c} 
		\hline
		Surface & Space & $p_\mathrm{s}$ & $k_\textrm{s}$ & $p$ & $\tau^2$   \\
		\hline
	 	 Pleiades & 	Linear & 	-0.1 & 	200.0 & 	2.4 & 	2749 \\ 
	 	   & 	Log & 	-0.1 & 	200.0 & 	2.3 & 2779 \\
		\hline
	 	 Praesepe & 	Linear & 	-0.1 & 	100.0 & 	2.2 & 4012\\	
	 	  & 	Log & 	0.1 & 	200.0 & 	2.1 & 3999\\	
	 	   \hline
	 	   Total&\textbf{Linear} &  \textbf{	-0.1} & \textbf{	100.0} & 	\textbf{2.2} & \textbf{6853} \\ 
		 	  & Log  & 	0.1 & 	200.0 & 	2.1 &  	6864 \\
		\hline
	\end{tabular}
\end{center}
\label{Tab:Matt2020}
\end{table}

Using Figure \ref{bestFits} as a guide, the poorer fit to Praesepe may be explained by the models' inability to fit both the low mass population of stars at longer rotation periods and the sequence of slow rotation at higher masses well. The Pleiades model has a somewhat uniform model density across all rotation periods up to a mass of about 0.8 $M_{\odot}$, above which the model density begins to increase as the model begins to converge. For Praesepe, this transition has occurred for more stars, reaching down to about 0.5 $M_{\odot}$. The model sequence occurs at much slower rotation than for the majority of observed data points. This misplaced high model density that does not correctly describe the high density of observed points gives rise to a much worse $\tau^2$ fit. The fact that a higher fraction of stars in Praesepe have transitioned to this sequence only exacerbates the mismatch in the fit.

\subsection{$\tau^2$ with different mass transformations}
\label{tau2differentmasstransforms}

To assess how sensitive $\tau^2$ is to different methods of mass transformations, we repeat the grid search exercise described in Section \ref{Matt2020Tau2Search}, but instead calculate a $\tau^2$ surface for just Praesepe using the mass dataset of \cite{Douglas2017}. The results are compared to those of Section \ref{Matt2020Tau2Search} where we used our own Praesepe mass dataset. The difference in the mass transformation methods are discussed in Section \ref{MassTransforms}.

The resulting $\tau^2$ surfaces are very similar to those shown in Figure \ref{bestFitSurfaces}, where we had used our own Praesepe dataset, in that each surface shows a low $\tau^2$ valley and a sharp increase in $\tau^2$ for $k_\textrm{s}$ $\leq$ 0. Table \ref{Tab:Matt2020Douglas} shows the differences in $\tau^2$ and the best-fit parameters for each of the two Praesepe datasets, where the results from the previous Section have been re-tabulated from Table \ref{Tab:Matt2020} for ease of comparison. In this incidence, the $\tau^2$ values and surfaces for Praesepe are directly comparable, since in each case the model set-up remains unchanged and the datasets contain the same number of stars. 

The best-fit parameters obtained from linear rotation space for each surface differ only in $p$, which changes from 2.2 to 2.0.  The corresponding $\tau^2$ values changed from 4012 (this work) to 4033 (literature dataset). This difference indicates that $\tau^2$ can detect a real difference between the datasets. 

It was a concern that in this particular case the difference in $\tau^2$, at 21, was not large compared to the differences between pixels in the parameter space. This is an indication that the $\tau^2$ parameter space is not sampled finely enough. A finer sampling of the parameters (not shown) was performed in linear space to obtain a better reflection of the true value of the minimum $\tau^2$. Using the finer grid, the best-fit parameters became $k_\mathrm{s}$\,=\,150, $p_\mathrm{s}$\,=\,0.0, p\,=\,$2.1$, at $\tau^2$\,=\,3987 for our own dataset, and $k_\mathrm{s}$\,=\,125, $p_\mathrm{s}$\,=\,$-0.05$, $p$\,=\,$2.1$, at $\tau^2$=4031 for the literature dataset. The difference in $\tau^2$ between pixels close to the minimum (which reflects the uncertainty within which we can determine $\tau^2$) is around 8. The difference in $\tau^2$ of 21 between datasets was greater than the uncertainty, showing that $\tau^2$ is able to detect a difference between the two datasets, with our data being a better fit.

Any statistical preference for one dataset over the other is likely sensitive to the model age, and would require thorough further investigation. Our aim however is not to state a preference for one dataset over another, but to demonstrate the sensitivity of the $\tau^2$ method, its ability to detect differences between the mass transformation methods, and to show that the choice of the latter can affect the resulting best-fit parameters.

\begin{table}
\caption{The best-fit stellar wind torque parameters and associated $\tau^2$ obtained in both log and linear rotation period space for a literature Praesepe dataset  \citep{Douglas2017} compared with the dataset obtained in this work.}

\begin{center}
	\begin{tabular}{ l|c|c|c|c|c } 
		\hline
		Surface & Space & $p_\mathrm{s}$ & $k_\mathrm{s}$ & $p$ & $\tau^2$   \\
		\hline
		\hline

	     \hline
	    	Praesepe   & 	Linear     & -0.1	 & 	100.0 & 2.0	 & 4033 \\ 
 		 (literature) & 	Log     & 	-0.1 & 	100.0 & 	2.1 & 	4002 \\ 
 		\hline
 		 Praesepe& 	Linear & 	-0.1 & 	100.0 & 	2.2 & 4012 \\
	 	  (this work)  & 	Log & 	0.1 & 	200.0 & 	2.1 & 	3999 \\ 

         \hline
         
	\end{tabular}
\end{center}
\label{Tab:Matt2020Douglas}
\end{table}

\begin{figure}[!htb]
\centering
      \includegraphics[trim=0.0 1. 0. 0., clip, width=1.\linewidth]{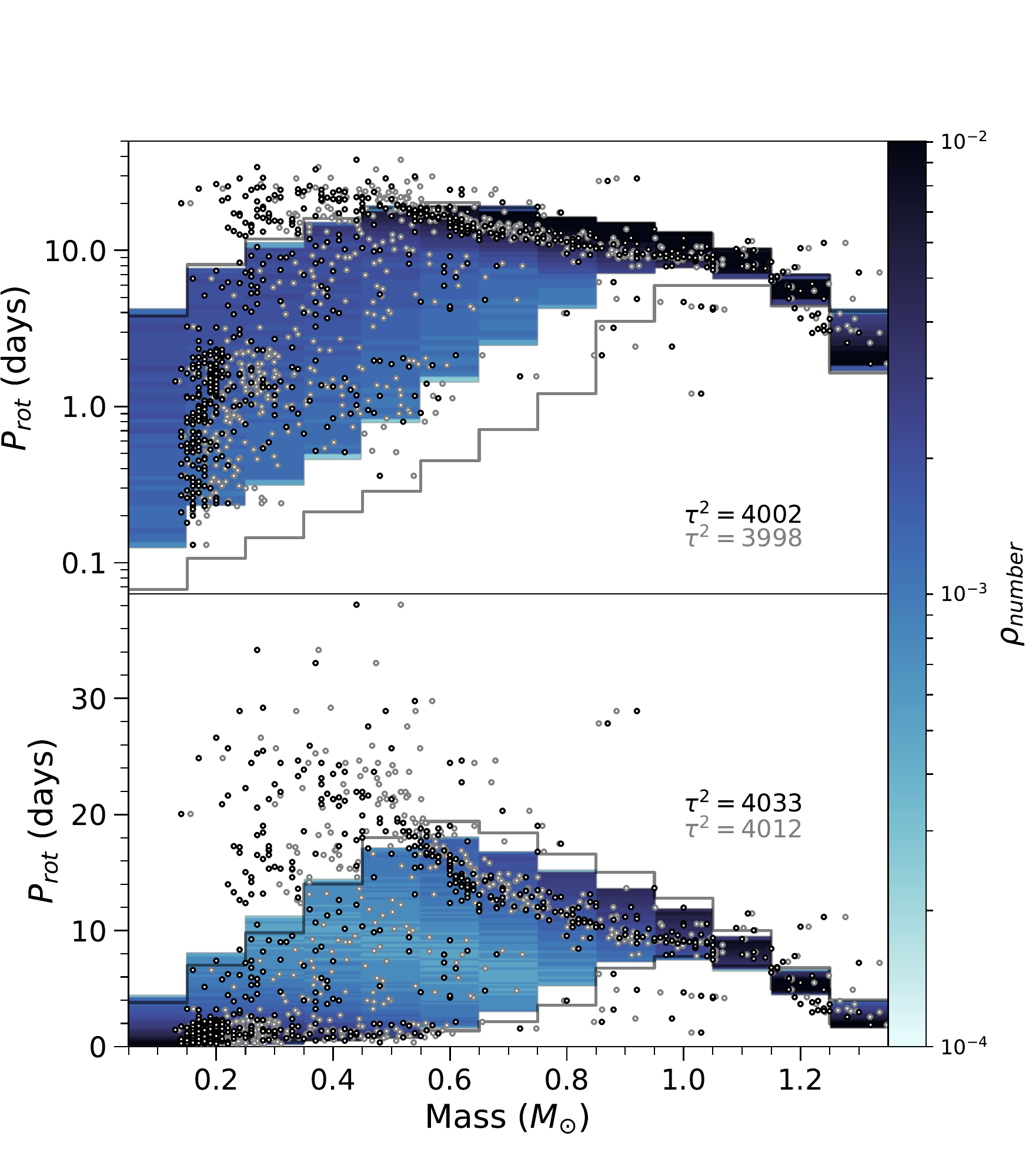}
     \caption{Period-mass diagrams showing the literature mass dataset for Praesepe in black, in both log period (top) and linear period (bottom). The model tuned to this dataset in log space (with best-fit values of $p_\mathrm{s}$\,=\,$-$0.1, $k_\textrm{s}$\,=\,100 and $p$\,=\,2.1, as shown in row 2 of Table \ref{Tab:Matt2020Douglas}) is shown as a density distribution in the upper panel, whilst the model fit in linear space (which differs only with a $p$ value of 2.0) is shown in the lower panel. The model number density $\rho_\mathrm{number}$ is represented by the colourbar, where darker regions of blue denote where the model is most probable (note that $\rho_\mathrm{number}$\,=\,$A_{ij}\rho'$). Each panel shows the corresponding $\tau^2$ goodness of fit value of each fit. Our own Praesepe mass dataset is shown as grey points for comparison, and we also show the extent of the models obtained using this dataset and the corresponding $\tau^2$ in grey. The parameters used to generate these models are in rows 3 (for log space) and 4 (for linear space) of Table \ref{Tab:Matt2020Douglas}.}
    \label{bestFitsDouglas}
\end{figure}

The best-fit parameters of the literature dataset were used to generate new model period-mass distributions, shown in Figure \ref{bestFitsDouglas}. The top panel shows the best-fit model in log space, generated using the best-fit log parameters. The bottom panel shows the model generated using the best-fit linear parameters in linear space. The Praesepe dataset of \cite{Douglas2017} used to tune these new models is shown in black. For ease of comparison, the outline of the best-fit models found in each space using our own dataset for Praesepe is shown in grey, along with our dataset (grey points). The parameters used to generate these four models are in Table \ref{Tab:Matt2020Douglas}. The resulting best-fit model distribution for each dataset is detectably different between mass transforms, showing that $\tau^2$ is sensitive to different mass transformation methods. Different datasets lead to a systematic shift in the best-fit model parameters. The uncertainty in stellar masses therefore directly translates into an uncertainty in our model parameters.

The two datasets give perhaps a less drastic difference in the $\tau^2$ fits and in the distribution of the model on the PMD than one might expect for a couple of reasons. First is that the mass resolution of our model is quite coarse at 0.1 $M_{\odot}$ in width, and secondly, as seen from Figure \ref{MassTransforms}, it is only the lowest mass stars that are most susceptible to change between different mass transform methods are below around 0.6 $M_{\odot}$.

\subsection{$\tau^2$ with different initial conditions }
\label{UpperScoInitCond}

There is significant structure in the observed spin distributions of the Pleiades and Praesepe which the models lack, particularly for the rapidly rotating stars at lowest masses. Some of this structure appears to be inherent from very early ages. The pre-main sequence cluster Upper Sco, at an age of about 8 Myr, already shows significant structure in its period-mass diagram, and at its young age, the stellar wind torques have not had much time to act on the rotation rates of these stars. Much of this structure can be introduced into the models by using a young cluster as an initial condition, as discussed in \cite{Somers2017}. Based on these previous works, we implement Upper Sco as an initial condition in our models in Section \ref{impInitCond} to show the effects of different initial conditions on the $\tau^2$ fit. To determine how more realistic initial conditions will affect the $\tau^2$ fits, we require a new approach for implementing the initial conditions. 

\subsubsection{Implementation of Initial Conditions as a Weighting of Model Stars}
\label{impInitCond}

In this Section, we describe our method of assigning weightings to model stars, which allows us to model more complicated initial spin distributions. Thus far, we have adopted a `Tophat', a uniform distribution in mass and log rotation period. However, any probability distribution function which represents the desired distribution of rotation rates can be applied. Here we show how we implement one such function representing the observed cluster spin distribution of Upper Sco. We can simply evolve the observed distribution of Upper Sco forwards in time, but then (1) the coverage of rotation periods across mass is uneven and (2) there are too few data points to construct reliable probability densities on the PMD. Instead, we construct a function that gives a probability density function (PDF) based on the observed dataset.

We begin with an initial broad Tophat spin distribution, with 500 stars in each mass bin with initial periods sampled between 0.1 and 20 days. This is evolved from the initial age (8 Myr in this case, the age we assume for Upper Sco) to some final time. By assigning each model star a weighting $w(M,P_{initial})$ from a PDF, we can then simulate any initial period distribution in post-processing. We define these weightings such that they sum to unity across the entire PMD. 

The initial range in rotation periods of the broad Tophat must be wide enough to encapsulate the desired spin distribution. This leads to the problem of some stars rotating faster than break-up speeds, but the advantage is that one only needs to set up an initial synthetic cluster and then allow it to evolve \textit{once}, and any desired `initial' conditions can be applied afterwards. 

At a given age, the model density $\rho(M,P)$ in a bin is calculated as
\begin{equation}
\label{binningWeightedModelDensities}
\rho(M,P) = \frac{\sum w(M,P_\mathrm{initial})}{A(M,P)},
\end{equation}
where $\sum w(M,P_\mathrm{initial})$ is the sum of the weights of stars within the bin at $(M,P)$, which is equal to the total normalised probability of the spin model at that location. $ A(M,P)$ is the area of the bin in units of $M_{\odot} \textrm{day}$. 

To obtain the PDF of Upper Sco, and hence the weightings of each model star, we first binned the observed data into 0.1 $M_{\odot}$ bins to mirror our spin model resolution. For each mass bin, a Kernel Density Estimate (KDE) is then generated in log space to obtain a smooth PDF. Our KDE method approximates each observed point as a Gaussian (we used a bandwidth of 0.3 days). The individual Gaussians are then summed to give a PDF for each mass. Since these functions integrate over all periods to give unity, the problem of having fewer stars in each mass bin is solved.

From these probability density functions, weightings for each model star are obtained using 

\begin{equation}
w(M,P_\mathrm{initial})= \frac{1}{13} \rho_\mathrm{KDE}(M,P_\mathrm{initial}) \Delta P,
\end{equation}

\noindent where $\rho_\mathrm{KDE}(M, P_\mathrm{initial})$ is the probability density of the KDE at the mass and period of the model star. We multiply this by the log period spacing $\Delta P$ of the Tophat spin model. The factor of $1/13$ arises due to the total probability of each KDE being unity, and since there are 13 KDEs, the total model probability across the PMD must still normalise to 1. 

Since the PDF is a continuous distribution spanning a much wider range in log period than that of our initial broad Tophat, any probability missed in the wings is accounted for by re-normalising the distribution to unity. 

\begin{figure}[!htb]
\centering
 \includegraphics[trim=4. 1. 4. 1., clip, width=.9\linewidth]{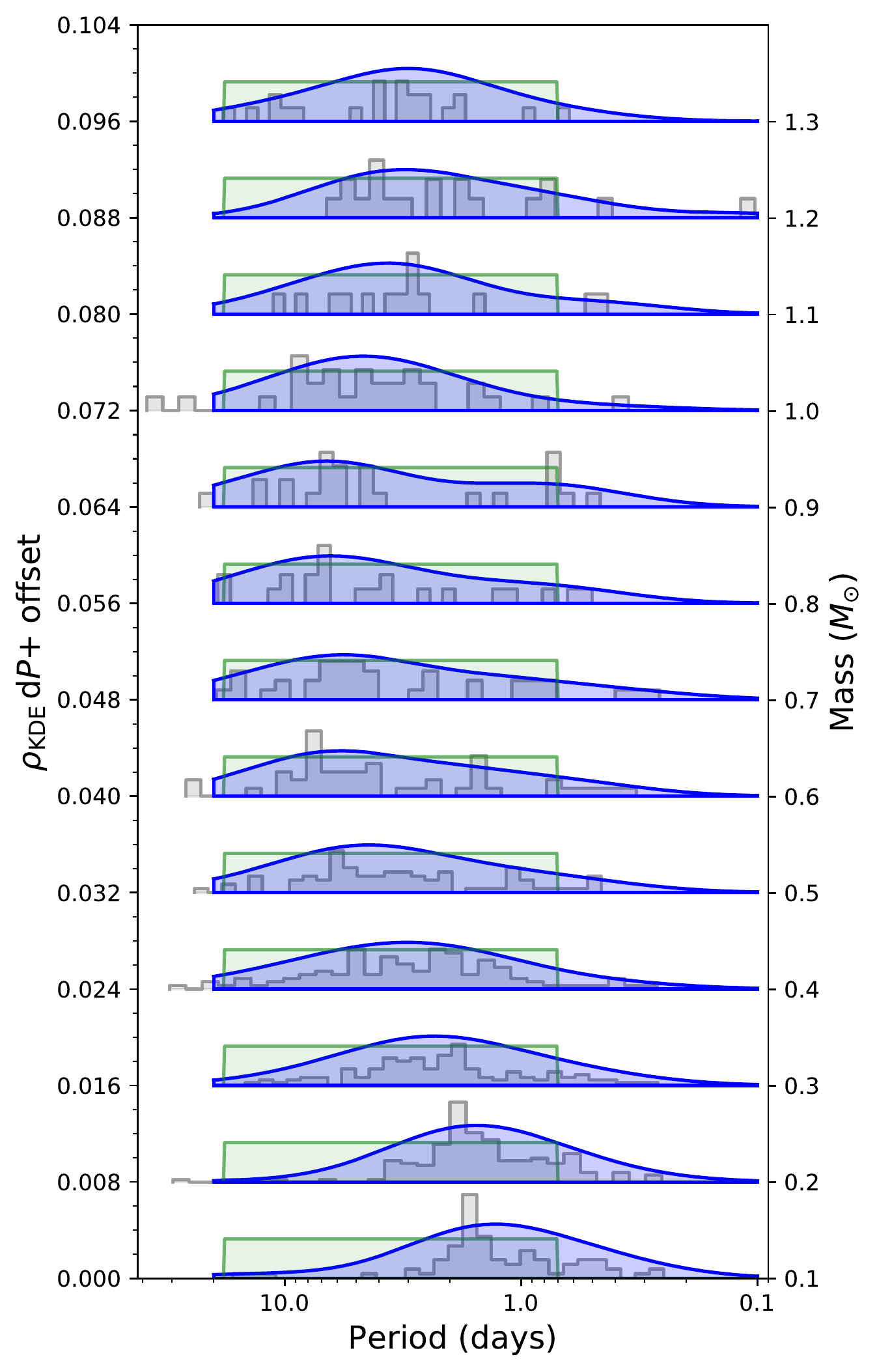}
     \caption{Initial probability distributions of spin periods for each mass bin, going from 0.1 on the bottom of the plot to increasing mass bins towards the top. Histograms of the Upper Sco dataset, each normalised to 1/2,  are shown in grey. The Tophat initial conditions, shown in green, are uniform in log period and do not vary with mass. Blue is the resulting smoothed KDE distributions of the observed Upper Sco dataset. The probabilities of each rotation period of each mass bin correspond to the height of each distribution. These distributions $f(M_\mathrm{initial},P)$ are used as initial conditions in the model.}
\label{KDEs}
\end{figure}

The resulting weightings of each of these stars are shown in Figure \ref{KDEs}. The two model initial conditions of Upper Sco and the Tophat used throughout this paper are shown. The Upper Sco distribution has been initialised at 8 Myr and the Tophat at 5 Myr. A Tophat evolved to 8 Myr would be shifted to faster rotation but would not contain any of the significant structure seen in the Upper Sco probability distributions, which will act to introduce significant structure to the spin models. The probability density of the spin model on the period-mass plane in each model gridcell is then calculated using Equation \ref{binningWeightedModelDensities}.

\subsubsection{Effect of Initial Conditions on $\tau^2$ Fit to Data}

Following a similar approach to the previous two Sections, we vary the wind parameters of $k_\textrm{s}$, $p_\mathrm{s}$ and $p$ methodically in a grid search to minimise $\tau^2$. The resulting $\tau^2$ surfaces corresponding to the best-fit $p$ values for the Pleiades, Praesepe and the total $\tau^2$ show the same general trends as those obtained previously. The best-fit parameters for each of the surfaces are listed in Table \ref{Tab:tau2results}. The $\tau^2$ values calculated with Upper Sco initial conditions are consistently lower in all cases by at least 90 than those in Tables \ref{Tab:Matt2020} and \ref{Tab:Matt2020Douglas}. This statistically verifies that using Upper Sco is a better choice than a flat Tophat initial condition, which is in agreement with \cite{Somers2017}.

When compared with the results in Sections \ref{Matt2020Tau2Search} and \ref{tau2differentmasstransforms}, the best-fit parameters in Table \ref{Tab:tau2results} vary again slightly, showing the ability of $\tau^2$ to detect a significant difference between initial conditions. They do however remain in broad agreement with previous Sections, with the total $\tau^2$ requiring a $p$ of 2.0, a $p_\mathrm{s}$ of -0.1 and a $k_\textrm{s}$ of 100. 

This being said, since the variation in best-fit parameters is only very slight when using different initial conditions, the simple Tophat initial condition is sufficient for probing the best-fit parameters in the torque law. Its downfall is that it will not give rise to desired structure in the period-mass plane, leading to a poorer $\tau^2$ fit.

\begin{figure*}
\centering 
\includegraphics[trim=2cm 0.1cm 0cm 1cm,clip,width=0.9\linewidth]{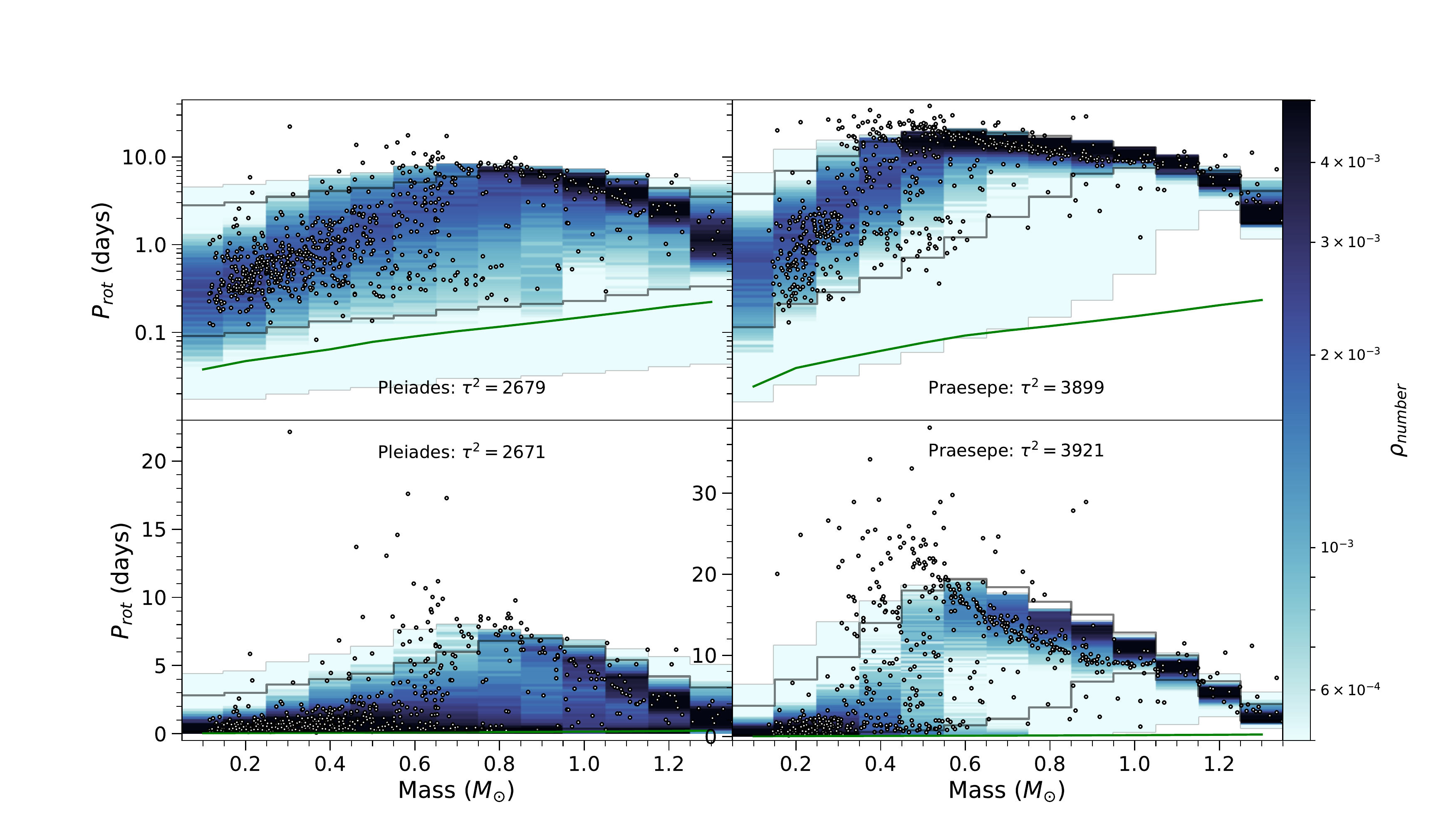}
\caption{Period-mass diagrams as in described in Figure \ref{bestFits}. The model distribution shown here instead uses Upper Sco as an initial condition with best-fit values from the total $\tau^2$ results shown in Table \ref{Tab:tau2results} ($p_\mathrm{s}$\,=\,$-$0.1, $k_\textrm{s}$\,=\,100 and $p$\,=\,2.0). The rotation period corresponding to the break up limit is shown in each plot by the green boundary. The extent of the corresponding best-fit model shown in Figure \ref{bestFits} is shown in grey for comparison.}
\label{bestFitsUpperSco}
\end{figure*}

The best-fit parameters of the total linear $\tau^2$ surface (bold row in Table \ref{Tab:tau2results}) is used to obtain the model probability density on the period-mass plane. This is shown in Figure \ref{bestFitsUpperSco}, where the model evolved to the Pleiades and Praesepe age is shown on the right and left respectively, in both log and linear space (top and bottom rows). The extent of the best-fit model shown in Figure \ref{bestFits} is shown in grey for comparison. 
The initial conditions of Upper Sco show an immediate improvement on the PMD. The envelope of the resulting model is significantly wider in the rotation period dimension (the extent of the model includes the light blue), meaning that the model is able to describe more of the faster rotators. In addition, the peak of the model density in each mass bin within this envelope is centred around the majority of the data, while still able to largely describe some of the slower rotators. This model does not place unnecessary model probability density above the bulk of the slower rotating low mass stars, where a Tophat would give a mostly flat probability across all rotation periods. 
However, much of the model predicts stars that spin too quickly to be physical, represented in Figure \ref{bestFitsUpperSco} by the green boundary. This is a natural consequence of using a wide initial condition produced by the KDEs.

While using Upper Sco as an initial condition shows obvious improvements, both from its distribution on the period-mass plane shown in Figure \ref{bestFitsUpperSco} and from the $\tau^2$ values listed in Table \ref{Tab:tau2results}, the model still suffers from the same issues as the models in the previous two Sections. The slowest rotators at lowest masses still cannot be properly fit, and the converged sequence  is predicted to spin down too quickly compared to the bulk of the observed data.

\begin{table}
\caption{Similar to Table \ref{Tab:Matt2020}, the best-fit wind parameters and $\tau^2$ for each cluster, and the combined $\tau^2$ results obtained when using Upper Sco as an initial condition to the spin models. The row in bold corresponds to the parameters used to generate the models shown in Figure \ref{bestFitsUpperSco}.}
\begin{center}
	\begin{tabular}{ c|c|c|c|c|c } 
\hline
Surface & Space &  $p_\mathrm{s}$ & $k_\textrm{s}$ & $p$ & $\tau^2$   \\
\hline
\hline
 	  	Pleiades & 	Linear & 	-0.2 & 	100.0 & 	2.3 & 2615 \\
 	  	 & 	Log & 	-0.2 & 	100.0 & 	2.3 & 	2612 \\

	\hline
 	 	Praesepe & 	Linear & 	-0.1 & 	100.0 & 	2.0 & 		3921 \\ 
 		 	 & 	Log & 	-0.1 & 	100.0 & 	2.0 & 	3899 \\ 
 		 	\hline
	 	Total	  & \textbf{Linear} &  	\textbf{-0.1} & \textbf{	100.0 }& 	\textbf{2.0} &  \textbf{	6593} \\
	 	 	  & Log  & 	-0.1 & 	100.0 & 	2.0 &   	6578  \\
	\hline
	\end{tabular}
\end{center}
\label{Tab:tau2results}
\end{table}

\section{Resulting Torque Laws}
\label{ResultingTorqueLaws}

Of all the models analysed in this paper, we find the best-fitting to be one where the observed spin distribution of Upper Sco is implemented as an initial condition (see Section \ref{UpperScoInitCond}). The $\tau^2$ fits of this model to the Pleiades, Praesepe, and both clusters simultaneously consistently give the lowest $\tau^2$ values (see Table \ref{Tab:tau2results}) of all our model setups (Tables \ref{Tab:Matt2020} and \ref{Tab:Matt2020Douglas}).

\begin{figure*}
\centering 
\includegraphics[width=0.9\linewidth]{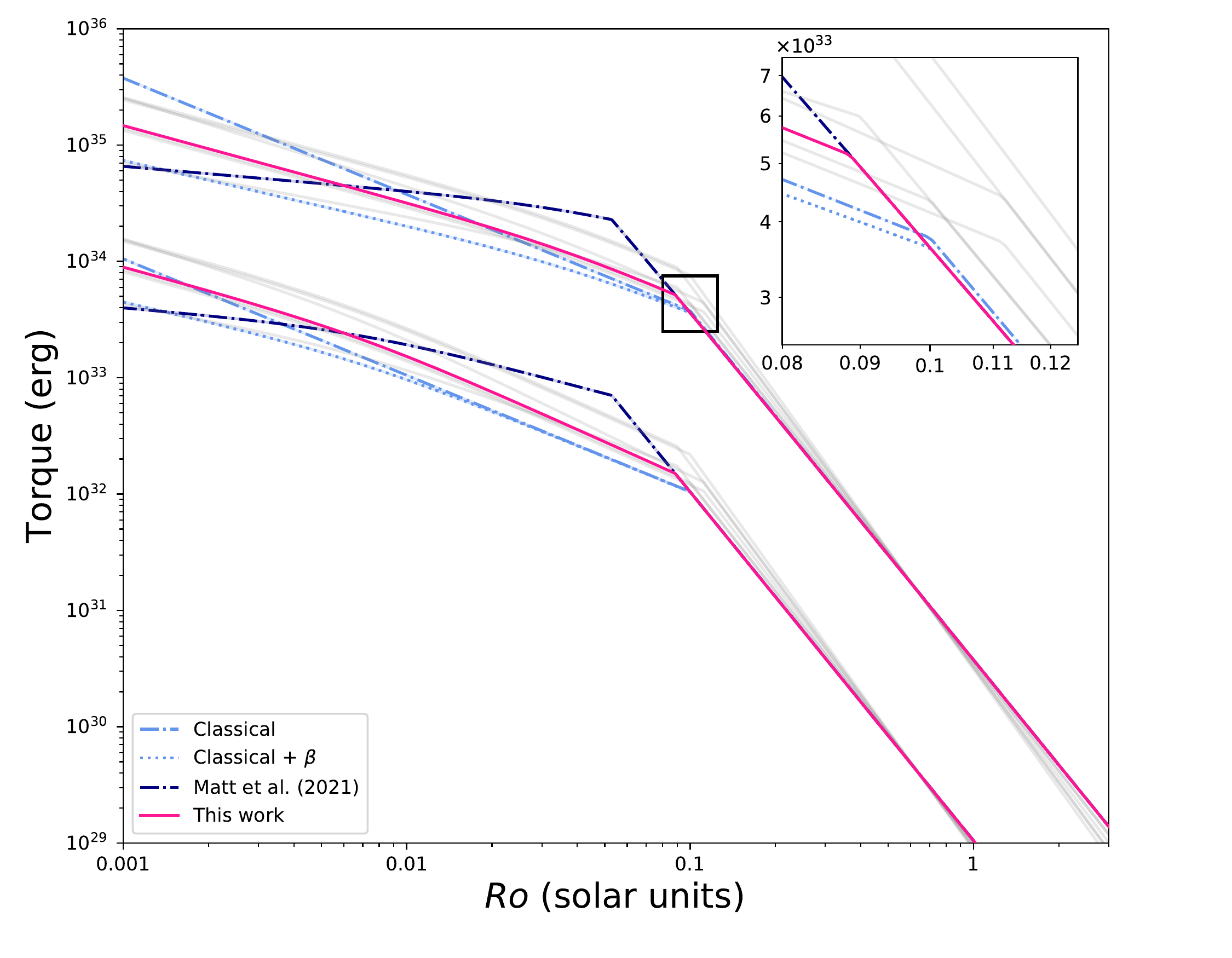}
\caption{The total stellar wind torques as a function of Rossby number (Equation \ref{torqueLaw}) for a variety of cases. The upper collection of lines corresponds to a 1 $M_{\odot}$ star; the lower collection corresponds to a 0.5  $M_{\odot}$ star. The panel in the upper right is a closer view of the break in the power laws of the 1 $M_{\odot}$ star, indicated by the square. The dark blue dot-dashed lines are our Standard model scalings for the torque in the saturated and unsaturated regimes. The light blue dashed and dot-dashed are scalings from \citealt{Matt2015} (the Classical model). The variety of grey dashed lines are the torque laws found in Section \ref{Results} of this paper. The pink line is our preferred torque law which is overall the best-fit, which is found using our datasets for the Pleiades and Praesepe, and initial conditions based on the observed spin distribution of Upper Sco. For each scenario, the variation in best-fit parameters (and hence the torques) demonstrates the ability of the $\tau^2$ technique to detect differences in the model set-ups and datasets, while also showing that the best-fit parameters found in all cases vary only slightly and in general agree very well. }
\label{torqueRossbyDependence}
\end{figure*}

Figure \ref{torqueRossbyDependence} shows the stellar wind torque of this best-fit (and hence our preferred) model prominently in pink for a 1 and 0.5 $M_{\odot}$ star. Also shown in grey are the remaining torque laws which have been found in the linear period-mass plane (see Section 
\ref{Results}). We prefer the linear space since we find in Section \ref{testingTau2} that it gives a consistent minimum even for a low resolution of model densities on the PMD. The slight variation of all the torque laws found in this paper (grey and pink) is testament to the sensitivity of $\tau^2$, and its ability to detect differences in each case. Remarkably, despite the variation between torque laws, the best-fit parameters agree to a narrow range of values.

Also shown in Figure \ref{torqueRossbyDependence} are the Classical model, the Classical model with an included $\beta$ term, and our `Standard model', which we define here as having a $\beta$ described by Equation \ref{beta}, $k_\mathrm{s}$\,=\,450, $p_\mathrm{s}$\,=\,0.2 and $p$\,=\,2. We have chosen these parameters such that the model does better at reproducing the observed `gap' in rotation rates for young clusters such as Praesepe, but it is important to note that the model does so without regards to how it fits the observed features in the PMD as a whole. Our holistic $\tau^2$ technique instead finds parameters that are the best compromise for fitting all observed features simultaneously. It is clear that each of the best-fit torque laws found from the $\tau^2$ approach is very close to those of \cite{Matt2015} and the Standard model. In all cases, the critical Rossby number whereby the torques change from the saturated to the unsaturated regime is very similar to that of \cite{Matt2015}. The saturated regime torques tend to be neither as shallow as that predicted by \cite{Matt2015}, nor as steep as that of the Standard model, instead falling between the two.

We have found in our preferred model the set of parameters and initial conditions that best describe the data with our torque law formalism. This best-fit torque law does not differ drastically from those of \cite{Matt2015} and the Standard model. Our statistical analysis does not find a requirement for inefficient torques at fast rotation as has been proposed by a few recent works \citep[e.g.][]{Brown2014,Garraffo2018}, despite being able to accommodate for this scenario with a significantly positive $p_{\mathrm{s}}$ value in our torque law. Moreover, the resulting model distribution on the PMD shows that our preferred model fits the slowly rotating population of M dwarfs well (see Figure \ref{bestFitsUpperSco}), which is the population of stars that is purported to require inefficient braking at low Rossby numbers.

However, the distribution of even our preferred model on the period-mass plane shows that the models require further improvement. The model converging sequence at higher masses predicts stars to rotate much more slowly than is observed, which appears to be related to the phenomenon of the temporary stalling of stellar spin-down recently discussed by \cite{Curtis2020}. Additionally, the model does not describe the low-mass, slow-rotators well (see e.g. lower-right panel of Figure \ref{bestFitsUpperSco}). To better fit the data, our rotational evolution models require improvement. For example, they could include more complex stellar wind torque laws or core-envelope decoupling, as in \cite{Spada2020}. While our $\tau^2$ technique is well-suited to determine best-fit parameters for these more complex models, doing so is outside the scope of this paper.

Finally, throughout this paper, our synthetic clusters use solar metallicity. Although the clusters we have used in our fitting are both super-solar \citep{Soderblom2009,Dorazi2020}, we have assumed the difference in metallicity between the Pleiades and Praesepe is sufficiently small that our fits have not been affected. It has recently been found that metallicity can alter the rotational evolution of stars \citep{Amard2020}. Including the effects of metallicity could thus impact on our resulting torque laws.

\section{Conclusion}
\label{Conclusion}

We have developed the maximum likelihood fitting statistic, $\tau^2$, to operate in the two-dimensional period-mass plane. For a given spin model, the $\tau^2$ statistic returns the goodness of fit to an observed dataset. The main findings of this paper are as follows.

\begin{enumerate}[leftmargin=+0.2in]
\item In Section \ref{testingTau2}, we demonstrated in Figure \ref{gyroPlot} the technique's ability to produce for the first time a statistical best-fitting Gyrochronology age which relies not on a single one-dimensional sequence Gyrochrone, but utilises both a physically motivated rotational evolution model and an observed spin distribution in two-dimensional space to obtain a fit. This result shows the potential of $\tau^2$ in future Gyrochronology age measurements for populations of stars. 
\item We additionally showed that, for known ages, the $\tau^2$ technique can be used to test the stellar wind torque prescription. Best-fit values for the three torque parameters $k_\textrm{s}$, $p_\mathrm{s}$ and $p$ were found and compared for three different cases.
    \begin{enumerate}[leftmargin=+0.2in]
        \item The Pleiades and Praesepe in Section \ref{Matt2020Tau2Search} each required slightly different best-fit torque parameters (see Table \ref{Tab:Matt2020}), which we interpret to be either an artefact of the models being unable to fit all features in the period-mass diagram, or as possible inter-cluster differences. In addition, Praesepe was found to be a more stringent constraint for the torque parameters than the Pleiades.
        \item Different mass transforms in Section \ref{tau2differentmasstransforms} gave rise to different best-fit parameters. We thus found that $\tau^2$ can detect differences between mass transformations. These results are listed in Table \ref{Tab:Matt2020Douglas}.
        \item The implementation of the more structured initial conditions of Upper Sco (rather than a uniform distribution in period and mass) in Section \ref{UpperScoInitCond} gave the best overall $\tau^2$ fit, with the resulting model spin distribution shown in Figure \ref{bestFitsUpperSco}. The best-fit parameters did not change drastically, showing that although more detailed initial conditions can improve the fit of the model, they cannot compensate for any flaws in the stellar wind torque law. 
    \end{enumerate}
\item We found that best-fit parameters found for $k_s$, $p_s$ and $p$ varied slightly between each of the three scenarios we tested in Section \ref{ResultingTorqueLaws}. These subtle differences highlight the sensitivity of the $\tau^2$ method, but the broad results - determined by fitting both the Pleiades and Praesepe - point to a relatively narrow range for these parameters. The values for the torque parameters are found most consistently to be $p_\mathrm{s}$\,=\,$-0.1$, $k_\textrm{s}$\,=\,$100$. The value for $p$ varies on a case-by-case basis but is between 2.2 and 2.0. In all cases, the resulting stellar wind torque laws fall near the torque laws of \cite{Matt2015} and our Standard model, as shown in Figure \ref{torqueRossbyDependence}. 

\item We found no evidence for extremely weak braking at low Rossby numbers, as proposed in a few recent works. Our broken-power law formalism is sensitive to this scenario, and can achieve this by having a significant positive saturated slope ($p_{\mathrm{s}}$). Despite covering an appropriately large parameter space, our statistically-determined torque laws do not predict inefficient torques at low Rossby numbers. Furthermore, our models describe the rapidly rotating M dwarfs well, which is the population of stars purported to require reduced braking. 

\end{enumerate}

\noindent Using the $\tau^2$ technique, we have found the best-possible fitting model which uses our torque law formalism, by varying the three parameters of $k_\textrm{s}$, $p_\mathrm{s}$ and $p$ and finding best-fit values for each. While this model, shown in Figure \ref{bestFitsUpperSco}, is statistically an improvement over the literature models of \cite{Matt2015} and our Standard model, it is clear that a more sophisticated torque law is still required to obtain a better fit in the period-mass plane. Such a model must better describe the shape of the converged sequence, which we have determined is not a problem of initial conditions but rather likely due to the torque law itself. The development of the $\tau^2$ technique shows promise, allowing us to find best-fit parameters for any torque law, and to test new torque laws, improving the spin evolution models and hence our understanding of stellar magnetism and winds.

\acknowledgments{ACKNOWLEDGEMENTS}

We would like to thank Louis Amard, Julia Roquette, Alessandro Lanzafame, Colin Johnstone, Marcel Agüeros and Stephanie Douglas for useful discussions. We also appreciate the comments given by the anonymous referee, and thank them for helping us to improve this paper. AAB and SPM would like to acknowledge funding from the European Research Council (ERC) under the European Union’s Horizon 2020 research and innovation program (grant agreement No. 682393 AWESoMeStars). AAB also acknowledges funding from the College of Engineering, Mathematics and Physical Sciences at the University of Exeter.

\bibliographystyle{apj}
\bibliography{AngiePaper}



\newpage
\end{document}